\pdfoutput=1
\documentclass[11pt]{article}

\usepackage[T1]{fontenc}
\usepackage{newtxtext,newtxmath}
\usepackage[raggedright]{titlesec}
\usepackage[margin=1in]{geometry}
\usepackage{setspace}
\usepackage{graphicx}
\usepackage{booktabs}
\usepackage{tabularx}
\usepackage{array}
\usepackage{enumitem}
\usepackage[font=small,labelfont=bf,labelsep=period,skip=6pt]{caption}
\usepackage{placeins}
\usepackage{microtype}
\usepackage{xurl}
\usepackage{xcolor}
\usepackage[colorlinks=true,linkcolor=black,citecolor=black,urlcolor=blue!50!black]{hyperref}  

\setlist{itemsep=0.2em,topsep=0.3em}
\newenvironment{references}%
  {\begin{list}{}{\setlength{\leftmargin}{0.4in}\setlength{\itemindent}{-0.4in}%
     \setlength{\itemsep}{0.25em}\setlength{\parsep}{0pt}}\small\raggedright}%
  {\end{list}}

\newenvironment{promptblock}{\begin{quote}\small\setlength{\parskip}{0.3em}}{\end{quote}}

\hypersetup{
  pdftitle={Harder Than the Task: From Technical to Organizational Difficulty in AI Benchmarking},
  pdfauthor={Victor Barres, Aden Barton, Nicolai J. Foss, Henrik M. Nielsen}
}

\begin{document}

\begin{center}
{\large\bfseries HARDER THAN THE TASK: \\ FROM TECHNICAL TO ORGANIZATIONAL DIFFICULTY IN AI BENCHMARKING\par}
\vspace{1.2em}
{\normalsize Victor Barres\textsuperscript{1}, Aden Barton\textsuperscript{1}, Nicolai J. Foss\textsuperscript{2}, Henrik M. Nielsen\textsuperscript{2}\par}
\vspace{0.4em}
{\small \textsuperscript{1} Mercor\par}
{\small \textsuperscript{2} Copenhagen Business School\par}
\end{center}

\begin{abstract}
\noindent AI benchmarks have made remarkable progress in measuring the ability of AI systems to perform increasingly demanding cognitive and technical tasks. Yet, many of these benchmarks treat technical tasks as the unit of analysis, while much economically valuable work takes place within organizations, where tasks are embedded in a system of relations. Within organizations, successful task performance depends not only on solving the technical demands of the task but also on navigating the interdependence of activities, distributed knowledge, differentiated authority, temporal constraints, conflicting interests, and changing environmental conditions while the task is underway. Accordingly, we argue that organizational task difficulty constitutes a distinct and insufficiently recognized dimension of AI capability, and that existing approaches to benchmarking provide an incomplete basis for assessing how effectively AI agents can perform real-world organizational work. By drawing on organization theory, we develop a taxonomy of the structure of work that characterizes the conditions that agents face in organizational work. Applying this taxonomy to analyze existing benchmarks, we find limited and uneven coverage of organizational dimensions, with fewer than one quarter of prominent benchmarks claiming any of them. We therefore argue that AI benchmarks should increasingly move beyond the evaluation of technically difficult tasks toward the systematic assessment of agents' ability to navigate the organizational conditions under which economically valuable work takes place.
\end{abstract}

\FloatBarrier
\section*{INTRODUCTION}
In the last few years, artificial intelligence models have made striking progress on benchmarks spanning many occupations, from coding to mathematics to law. These benchmarks are often taken as evidence of AI's ability to do entire jobs and therefore are sometimes seen as leading indicators of worker automation or displacement (\hyperlink{ref:mazeika2025-gatti}{Mazeika et al., 2025}). But most AI benchmarks evaluate atomic tasks, whereas most economically valuable work is performed in organizations. That is not to say that AI models cannot perform organizational work, but rather that few benchmarks measure such organizational capability; as a result, most benchmarks fail to measure the tissue that aggregates technical tasks into a job.

Task difficulty and the difficulty of performing work inside an organization are not the same thing. Tasks are here understood as the foundational, microscopic units of work that combine to form jobs, roles, occupations, and entire organizational workflows. Work in organizations takes the form of workflows, and organization theory has long recognized that organizing work involves securing the conditions this work requires: managing the dependencies between tasks (\hyperlink{ref:thompson1967}{Thompson, 1967}), drawing on knowledge dispersed across different actors (\hyperlink{ref:hayek1945}{Hayek, 1945}), and relying on hierarchies that route problems to the right problem-solvers (\hyperlink{ref:garicano2000}{Garicano, 2000}). A task may be technically straightforward once the relevant information, objective, and decision rights are specified, but such simplicity is often not present in organizations. Instead, much of the difficulty in organizational work arises from the surrounding structures rather than from the focal task in itself.

Consider a task in which a human resources employee, following a merger, needs to determine the service date for a group of employees. The underlying task may be quite trivial, but the correct answer could depend on information distributed across a handbook, a merger agreement, an external consultant, a benefits administrator, and former employees. Some sources may be in conflict, certain information may be confidential, and a looming deadline makes the order of interactions important. Completing the task well, then, depends not only on calculating the correct date but also on understanding how the task is embedded in a structure of information, authority, and interdependence.

We call the additional difficulty created by such organizational embedding \emph{organizational task difficulty}, in contrast to \emph{technical task difficulty}. Technical task difficulty arises from performing the focal transformation when the relevant objective, inputs, and authority to act are given. Organizational task difficulty arises when those inputs are not given and must be located, reconciled, or obtained from other activities and actors. The two can vary independently, where a technically demanding task may be organizationally simple, and a technically simple task can be organizationally demanding.

The advent of AI colleagues, such as Claude Tag (\hyperlink{ref:anthropic2026}{Anthropic, 2026}), and personal AI agents, such as Meta's Muse (\hyperlink{ref:meta2026}{Meta, 2026}), highlights how such agentic systems are increasingly capable of carrying out complex tasks on behalf of their users. Despite these advancements in AI capabilities, the economic effects observed so far have been more modest and uneven than what the pace of model development would suggest (\hyperlink{ref:acemoglu2025}{Acemoglu, 2025}; \hyperlink{ref:brynjolfsson2025-li}{Brynjolfsson et al., 2025}). One potential explanation among several for this gap is the distinction between technical and organizational difficulty. Once implemented in organizations, AI agents must navigate interdependencies, information, and authority, which are not necessarily solved by scaling technical capabilities. At the same time, AI agents have shown notable capabilities to coordinate among themselves, forming systems that function as organizations in their own right (\hyperlink{ref:arrieta2026-he}{Arrieta et al., 2026}), sometimes with few or no humans involved. The OpenAI--Hugging Face incident in July 2026 illustrates this ability to coordinate at scale. Around 1,200 agents coordinated through an unsanctioned message board in an attempt to game how they were scored, which led to the breach of Hugging Face's systems (\hyperlink{ref:metr2026}{METR, 2026}). This incident underscores the capabilities and risks introduced by AI agents, raising questions around the governance mechanisms when such AI capabilities are implemented in existing organizations.

Although AI agents are increasingly used in organizations, AI benchmarking lacks a vocabulary for describing the organizational conditions they face. Existing taxonomies provide rich frameworks for describing what work consists of. O*NET is one such taxonomy, providing a systematic classification of occupations (\hyperlink{ref:peterson2001-mumford}{Peterson et al., 2001}). Recent studies have applied it to assess how large language models (LLMs) may affect occupations and the broader economy (\hyperlink{ref:eloundou2024-manning}{Eloundou et al., 2024}; \hyperlink{ref:felten2023-raj}{Felten et al., 2023}; \hyperlink{ref:mertens2026-kuzee}{Mertens et al., 2026}; see also \hyperlink{ref:mercor2026}{Mercor, 2026}). Yet, AI benchmarking lacks a taxonomy comparable to O*NET for describing the organizational conditions surrounding task performance. Further, improvements in technical task capability need not be accompanied by equivalent improvements in the ability to navigate the interdependencies, authority, and information structures of organizations. As a result, without such a taxonomy, AI benchmarks have no systematic way to characterize the organizational situations they represent, identify important situations they omit, or locate the sources of model success and failure.

Our primary conceptual contribution addresses this gap. We develop a preliminary taxonomy that provides such a vocabulary for the classification of important organizational conditions surrounding task performance. Drawing on organization theory and organizational economics, we identify recurring features of organizational work, including the interdependence among activities, the distribution of information, the allocation of authority and decision rights, temporal dependencies, the alignment of interests, and changes in relevant conditions while work is underway.

We structure the organizational embedding of work into four analytically related layers: structure of work, coordination requirements, governance mechanisms, and performance outcomes. Our taxonomy focuses primarily on the structure of work, namely the organizational conditions facing actors, human and AI alike. At a higher level, we also highlight the coordination requirements created by the structure of work, the mechanisms for addressing them, and the resulting performance.

The taxonomy also creates a basis for asking what organizational conditions are actually captured by the benchmarks used to evaluate AI models. Aggregate benchmark scores can conceal systematic strengths and weaknesses of models when benchmarks combine heterogeneous organizational conditions without clearly identifying them. We therefore use the taxonomy to label the organizational coverage of existing AI benchmarks. In this analysis, we distinguish between the organizational features that are merely mentioned in benchmark descriptions, those that are claimed as part of the benchmark itself, and those that are measured separately in standalone results.

With this labeling, we find that organizational conditions are largely absent from the benchmarks the field pays most attention to. Fewer than a quarter of the 522 prominent benchmarks we tag (118, or 23 percent) claim any of the six dimensions that constitute work's organizational structure. And when organizational conditions appear, they are concentrated in a few forms. For example, authority most often takes the form of a written policy, information often requires reconciling documents, and work with colleagues is nearly absent. Such benchmarks may have realistic atomic tasks, which may be quite difficult technically, but the field has left organizational task difficulty largely unmeasured.

Our paper makes three contributions. First, we distinguish technical task difficulty from organizational task difficulty, providing a way to conceptualize why rapidly improving performance of AI models on technically specified benchmarks may translate imperfectly into performance inside organizations. Second, we forward a preliminary taxonomy for representing the organizational dimensions of work, which is intended to offer a common language for benchmark construction, coverage assessment, and comparison. Third, we apply the taxonomy to existing AI benchmarks to assess organizational coverage. Taken together, these contributions suggest shifting the unit of AI evaluation from the isolated task toward the task as it is performed within organizations.

\FloatBarrier
\section*{ORGANIZATIONAL DIFFICULTY AND THE LOGIC OF AI BENCHMARKING}

\subsection*{The sources of difficulty}
Much economically valuable work takes place within organizations. As \hyperlink{ref:coase1937}{Coase (1937)} noted, work is organized within firms, rather than among individuals trading through markets, because using the market to coordinate is costly for some activities, whereas within firms, work is instead directed by authority. Further, within firms, output depends on more than the execution of individual tasks. In this context, tasks are the foundational, microscopic units of work that combine to form jobs, roles, occupations, and workflows. Each task converts inputs into an output, such as goods and services (\hyperlink{ref:acemoglu2011-autor}{Acemoglu \& Autor, 2011}). Within organizations, tasks form parts of systems of differentiated activities with interdependencies (\hyperlink{ref:thompson1967}{Thompson, 1967}), and performance often depends on the actions, knowledge, and decisions of other actors. For example, a production engineer may require information that sits with the design department, require approvals from the finance department for procurement of parts, and depend on the engineer's manager for the ultimate decision to execute the plan. Difficulty thus arises not only from the focal task in itself, but also from the relationships that connect it to other activities and actors.

The first source of difficulty is what we call \emph{technical task difficulty}, which refers to the difficulty of performing the focal transformation when the relevant objective, inputs, and authority to act are given. It may arise from domain knowledge, reasoning requirements, the number of steps required, computational demands, or uncertainty embedded in the focal problem. Most of the existing AI benchmarks focus on tasks of this nature, such as with problems in coding, math, or reasoning.

The second source of difficulty, \emph{organizational task difficulty}, arises when successful performance of the focal task depends on relationships with other activities or actors. Much of what goes on in organizations is based on relationships with other actors, which makes this source of difficulty relational. This difficulty can arise, for instance, when relevant information sits elsewhere, when another actor controls a required input, when decision rights sit with somebody other than the focal actor, or when actors have different objectives. Organizational task difficulty thus reflects the relationship dependencies surrounding the task rather than technical demands of the focal transformation.

Organizational task difficulty is not just a reflection of the number of actors or interactions involved. Tasks with many actors may impose relatively modest coordination requirements if their activities are largely independent. Rather, it is the structure of dependencies connecting actors, activities, information, and decision rights that gives rise to such difficulty.

These dependencies do not create a single, uniform source of difficulty. Instead, the difficulty is closely connected to the task at hand and the organizational context in which it is executed. Simply collapsing these conditions into a uniform category of organizational complexity fails to capture how different demands arise, such as from distributed information or the divergent interests of actors. An actor may be perfectly capable of acquiring and validating different information sources yet unable to balance the competing interests of others within the organization.

\begin{figure}[htbp]
\renewcommand{\thefigure}{1}
\caption{Technical and organizational task difficulty}\label{fig:1}
\centering
\includegraphics[width=0.82\linewidth]{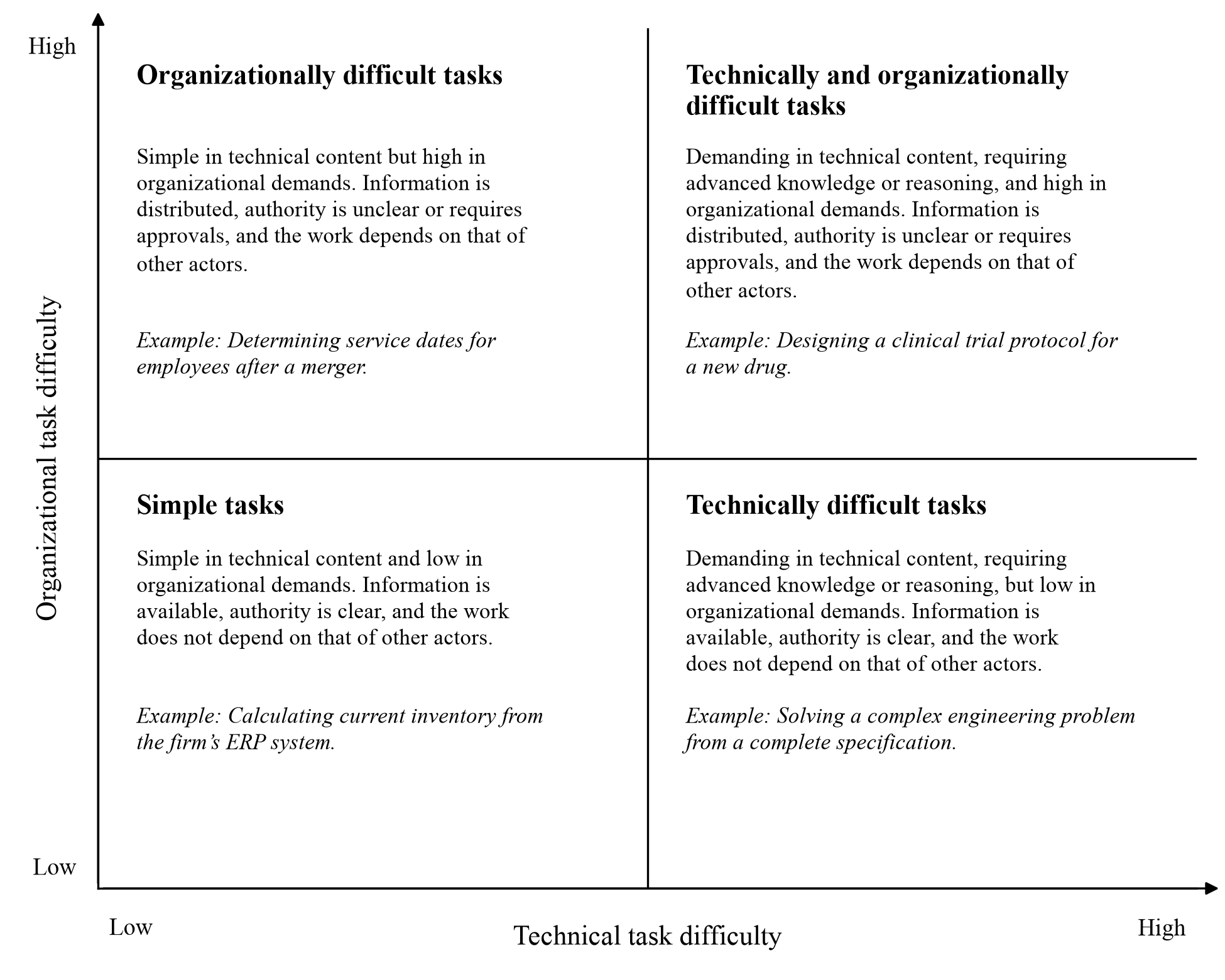}
\par\smallskip\begin{minipage}{\linewidth}\footnotesize\raggedright \emph{Note.} Illustrative. The axes are conceptual.\end{minipage}
\end{figure}

The distinction between technical and organizational task difficulty helps explain why certain relatively straightforward technical tasks may be difficult to complete within organizations. While both sources of difficulty are often present for the same task, the two can vary independently. A technically demanding task may be organizationally simple when the actor holds the relevant information, has the authority to act, and can perform the task without coordinating with others. Conversely, a task that is simple in its technical nature may be organizationally difficult when the information is distributed across multiple actors, when approvals must be obtained, or when successful completion depends on coordinating outputs with others. This independence implies that simplifying a technical task need not simplify the organizational setting in which it is performed, and vice versa. \hyperref[fig:1]{Figure 1} illustrates the four types of task that result.

The economics literature on technological change has similarly focused on tasks, although without explicitly distinguishing between their technical and organizational difficulty. In the task-based framework, jobs are seen as bundles of constituent tasks, and technologies can substitute for or complement particular tasks rather than occupations as a whole (\hyperlink{ref:acemoglu2011-autor}{Acemoglu \& Autor, 2011}; \hyperlink{ref:autor2003-levy}{Autor et al., 2003}). This task-based framework has been applied to AI, because its unit of analysis maps onto occupational taxonomies, such as O*NET, which provide relatively standardized descriptions of activities across different occupations (\hyperlink{ref:brynjolfsson2018-mitchell}{Brynjolfsson et al., 2018}; \hyperlink{ref:eloundou2024-manning}{Eloundou et al., 2024}; \hyperlink{ref:garicano2026-li}{Garicano et al., 2026}). The framework is not limited to technical work, as O*NET task descriptions often have organizational content, such as coordinating the work of others, but the tasks are necessarily described as discrete units separate from the organizational structure in which they are performed.

Although the task-based perspective has not provided a clean separation between the technical and organizational components of work, adjacent research has underlined the economic importance of the capabilities associated with operating in organizational contexts. At the individual level, research has documented the substantial and increasing returns that are associated with social skills in labor markets (\hyperlink{ref:deming2017}{Deming, 2017}; \hyperlink{ref:deming2018-kahn}{Deming \& Kahn, 2018}; \hyperlink{ref:edin2022-fredriksson}{Edin et al., 2022}; \hyperlink{ref:weinberger2014}{Weinberger, 2014}). Similarly, such social skills have also been shown to predict group performance at a level that is comparable to measured intelligence (\hyperlink{ref:weidmann2021-deming}{Weidmann \& Deming, 2021}). Social skills are not equivalent to organizational capability, but this research suggests that economically valuable work depends on more than technical proficiency.

Similarly, studies highlight that management practices account for substantial variation in firm-level productivity (\hyperlink{ref:bloom2007-vanreenen}{Bloom \& Van Reenen, 2007}), and that improvements in such practices can contribute to sizable productivity gains (\hyperlink{ref:bloom2013-eifert}{Bloom et al., 2013}). Related work emphasizes the economic importance of managerial quality for the productivity of workers and organizations (\hyperlink{ref:fenizia2022}{Fenizia, 2022}; \hyperlink{ref:lazear2015-shaw}{Lazear et al., 2015}). Taken together, the literature reinforces the broader point that productive performance depends not only on the ability to execute the focal task, but also on the way in which activities, information, and decisions are organized and coordinated.

The distinction between technical and organizational task difficulty becomes particularly relevant as AI systems continue to improve. AI models may keep advancing on technically specified tasks, but if there are not corresponding improvements for organizational tasks, the productivity gains from such model advancements may be limited. Existing research has already documented heterogeneity in the productivity effects of generative AI (\hyperlink{ref:brynjolfsson2025-li}{Brynjolfsson et al., 2025}), as well as outlined a ``jagged technological frontier'' of AI performance across tasks (\hyperlink{ref:dellacqua2026-mcfowland}{Dell'Acqua et al., 2026}). One possible source of this heterogeneity is the organizational setting in which tasks are performed.

\subsection*{The logic of AI benchmarking}
In contrast to the organizational difficulty described above, AI benchmarking has largely developed around a task-based conception of capability. AI benchmarks are tests used to score models. A benchmark specifies a problem, provides or makes available the information required to address it, and evaluates the model's response against some criterion of successful performance. Benchmarks serve two main functions, among others. First, they are a measurement tool, allowing researchers to compare systems on a common set of tasks (\hyperlink{ref:donoho2017}{Donoho, 2017}). Second, they can set direction. Once something can be measured, labs can hill-climb on it and make it a priority, meaning that benchmarks can shape what the field chooses to optimize (\hyperlink{ref:raji2021-bender}{Raji et al., 2021}). We focus on benchmarking, as opposed to AI evaluation more broadly, because the field has coordinated around benchmarks as a means of communicating model performance (\hyperlink{ref:hutchinson2022-rostamzadeh}{Hutchinson et al., 2022}; \hyperlink{ref:orr2024-kang}{Orr \& Kang, 2024}).

Benchmarks have changed as model capabilities have improved. The earliest benchmarks were often very narrow, built around single tasks such as image classification (\hyperlink{ref:deng2009-dong}{Deng et al., 2009}) or a small set of language-understanding tasks (\hyperlink{ref:wang2019-pruksachatkun}{Wang, Pruksachatkun, et al., 2019}; \hyperlink{ref:wang2019-singh}{Wang, Singh, et al., 2019}). As these tasks became saturated (\hyperlink{ref:kiela2021-bartolo}{Kiela et al., 2021}), the field moved toward broader, harder, and more realistic tests of knowledge and reasoning (\hyperlink{ref:hendrycks2021-burns}{Hendrycks et al., 2021}; \hyperlink{ref:phan2025-gatti}{Phan et al., 2025}; \hyperlink{ref:rein2024-hou}{Rein et al., 2024}). More recently, benchmarks have shifted from grading simple prompt responses toward measuring models as agents operating within environments, such as resolving code issues or scanning over websites (\hyperlink{ref:jimenez2024-yang}{Jimenez et al., 2024}; \hyperlink{ref:zhou2024-xu}{Zhou et al., 2024}). Similarly, benchmarks have moved toward tasks directly drawn from real professional work, in an attempt to measure how well models can contribute to productivity within a firm (\hyperlink{ref:kwa2025-west}{Kwa et al., 2025}; \hyperlink{ref:patwardhan2025-dias}{Patwardhan et al., 2025}).

Across this evolution, however, the underlying logic of the benchmark has stayed largely the same. Difficulty is typically varied through properties of the focal task: the knowledge it requires, the complexity of the reasoning involved, the number of steps needed to complete it, the length of the relevant context, or the technical quality of the resulting output. Improvements in benchmark performance are then interpreted as evidence of improvements in the model's capability to perform the corresponding class of tasks. This approach is well suited to measuring technical task difficulty.

However, the same design choices tend to abstract from organizational difficulty. In many benchmark settings, the objective is given, relevant information is either provided or retrievable, the agent possesses the authority needed to act, and the surrounding environment remains comparatively stable while the task is performed. These assumptions are ordinarily invisible because they are built into the format rather than stated in it. A task arrives as a prompt, and whatever is not in the prompt is either retrievable or irrelevant. Other actors may appear, particularly in newer interactive or multi-agent benchmarks, but their presence does not necessarily create organizational dependence. Often, the agent's own actions are the only source of change in the environment. Such a restriction is partly practical, since environments where other actors can alter the state of the world are harder to measure reliably and fairly. A common solution is to simulate these actors with LLMs, but whether LLM-simulated humans are valid stand-ins for real ones remains an open question (\hyperlink{ref:seshadri2026-cahyawijaya}{Seshadri et al., 2026}).

This model is not wrong so much as it is silent about the setting (and holding organizational conditions constant can in some situations be essential for identifying technical capability). The design of many benchmarks therefore holds the organizational setting constant at a particular value---the null value---and in doing so removes from measurement precisely the conditions that the previous section identified as the source of organizational difficulty.

\FloatBarrier
\section*{THE LAYERS OF ORGANIZATIONAL WORK}
The organizational setting surrounding a task can be described at several layers. Accordingly, we organize the organizational embedding of work into four analytically related layers across the structure of work, coordination requirements, governance mechanisms, and performance outcomes (\hyperref[tab:1]{Table 1}). The first layer, the \emph{structure of work}, concerns the interdependence of activities, the structures of information, authority, temporality, and incentives, as well as environmental dynamism. At the second layer sit \emph{coordination requirements}, which are generated by the structure of work. The third layer, \emph{governance mechanisms}, can help address these requirements. Finally, at the fourth layer, how well these requirements are met shapes the \emph{performance outcomes} at the organizational (or unit) level.

Consider a knowledge-intensive task within an organization where much of the relevant information is distributed across several actors (\hyperlink{ref:hayek1945}{Hayek, 1945}). Here, distributed information is a feature of the structure of work. This distributed nature of information creates a coordination requirement for the information to be acquired and reconciled before the task can be completed. The organization can accordingly put in place governance mechanisms, including communication channels, information systems (\hyperlink{ref:galbraith1974}{Galbraith, 1974}), or escalation procedures, in order to help meet that requirement. Finally, whether the actor executing the task manages to obtain and integrate the information correctly can be described as contributing to the performance outcome.

The taxonomy is intended as a guiding framework that can help the AI field draw on organization theory to evaluate how models navigate the organizational difficulty of tasks. It is not meant to be an exhaustive outline of organization theory but rather focuses on central ideas developed over the decades.

\begin{table}[htbp]
\renewcommand{\thetable}{1}
\caption{Four Layers of the Organizational Embedding of Work}\label{tab:1}
\centering\small
\begin{tabularx}{\linewidth}{>{\raggedright\arraybackslash}p{0.22\linewidth}>{\raggedright\arraybackslash}X>{\raggedright\arraybackslash}X}
\toprule
\textbf{Layer} & \textbf{Description} & \textbf{Examples} \\
\midrule
\textbf{1. Structure of work} & The organizational conditions surrounding the focal task & Task interdependence; information structure; authority structure; temporal structure; incentive structure; environmental dynamism \\
\addlinespace[0.6em]
\textbf{2. Coordination requirements} & The requirements placed on the connected actors based on the structural conditions & Acquisition and reconciliation of information; synchronization or mutual adjustment of inputs; decision routing and authorization; negotiation and alignment \\
\addlinespace[0.6em]
\textbf{3. Governance mechanisms} & The organizational means available to address the coordination requirements & Authority; communication; delegation; rules and standards; monitoring; sequencing; escalation \\
\addlinespace[0.6em]
\textbf{4. Performance outcomes} & The quality of focal output and whether organizational dependencies were handled successfully & Quality of the focal output; successful navigation of dependencies; proper organizational conduct (e.g., respect for decision rights, confidentiality, and task ownership); completion within resource and time constraints \\
\bottomrule
\end{tabularx}
\end{table}

\subsection*{Structure of work}
The structure of work focuses on the organizational conditions that the actor is faced with when executing a task, before the organizational mechanisms are brought into play. Drawing on extant research in organization design and organizational economics, we identify six dimensions: task interdependence, information structure, authority structure, temporal structure, incentive structure, and environmental dynamism (\hyperref[tab:2]{Table 2}).

\emph{Task interdependence} concerns how performance of the focal task depends on other activities.

\hyperlink{ref:thompson1967}{Thompson's (1967)} formulation of interdependence distinguishes between three forms, namely pooled, sequential, and reciprocal. \emph{Pooled interdependence} occurs when activities can be performed largely separately while still contributing to a common outcome. With \emph{sequential interdependence}, the activities are sequentially interconnected such that the output of one becomes an input to another. Finally, \emph{reciprocal interdependence} occurs when the outputs of each activity serve as inputs for the others, thereby requiring actors to adjust their activities to one another as work proceeds. As shown by empirical work, the different forms of interdependence are linked with different modes and intensities of coordination (\hyperlink{ref:vandeven1976-delbecq}{Van de Ven et al., 1976}). \hyperlink{ref:thompson1967}{Thompson (1967)} treats the three forms of interdependence as cumulative, in the sense that sequential interdependence incorporates pooled interdependence, while reciprocal interdependence incorporates both. In line with this, in our classification we distinguish tasks according to the most demanding form of interdependence they exhibit, as these forms imply different coordination requirements.

\emph{Information structure} concerns the distribution, accessibility, and character of knowledge needed to perform the focal task. Within organizations, the relevant knowledge may be concentrated with the focal actor or \emph{distributed} across people, units, and information systems (\hyperlink{ref:hayek1945}{Hayek, 1945}; \hyperlink{ref:march1958-simon}{March \& Simon, 1958}). Information within the organization may also be inconsistent or \emph{contradictory}, in the sense that different sources may present conflicting facts or opinions that the actor must resolve. Furthermore, some of the knowledge held by the actors within organizations is characterized by its \emph{tacitness}, where it cannot necessarily be put into words, but must instead be inferred from examples or practice (\hyperlink{ref:polanyi1966}{Polanyi, 1966}).

Importantly, \emph{organizational information dependence} arises when obtaining, interpreting, or validating task-relevant knowledge depends on other actors or on systems whose access, maintenance, or use is governed organizationally. This distinction separates organizational information dependencies from simple retrieval of information, as the information is not directly accessible to the actor, but hinges on other actors or systems controlled by the organization. Information-processing approaches to organization design have long treated the demands imposed by such information structures as central determinants of organizational arrangements (\hyperlink{ref:galbraith1974}{Galbraith, 1974}; \hyperlink{ref:tushman1978-nadler}{Tushman \& Nadler, 1978}).

\emph{Authority structure} concerns who can request, decide, approve, override, and oblige.

Within organizations, it is not a given that knowledge and authority reside with the same actor. Rather, an employee may possess the expertise required to identify and formulate an appropriate course of action, but not necessarily possess the right to implement it. Conversely, a manager may hold the formal authority over a decision, but not necessarily possess the most relevant knowledge. \emph{Formal authority} thus refers to the right to decide, typically derived from position or contract, while \emph{real authority} refers to the effective control over decisions that rests in part on holding relevant knowledge (\hyperlink{ref:aghion1997-tirole}{Aghion \& Tirole, 1997}). Within the field of organization theory, authority has long been treated as a means of coordinating specialized activities (\hyperlink{ref:simon1947}{Simon, 1947}), while organizational economics has also examined how decision rights can be allocated in response to the location and transfer costs of knowledge (\hyperlink{ref:jensen1992-meckling}{Jensen \& Meckling, 1992}).

\emph{Temporal structure} captures dependencies created by sequencing, deadlines, response latency, interruptions, and simultaneous streams of activity. Sequential interdependence of activities, as outlined earlier, already introduces a temporal ordering of tasks (\hyperlink{ref:thompson1967}{Thompson, 1967}), but work within organizations can also impose further temporal constraints. This is the case, for instance, when actors must wait for information or authorization from their managers, coordinate deadlines with other departments, or act within a time window to meet a deadline. Thus, task interdependence can introduce a sequential order of activities, while the temporal structure refers to when, at a given point in time, the task's actions must occur. Temporal structure is then not simply about how much time a task takes, but whether its successful completion depends on when other activities occur.

\emph{Incentive structure} concerns the alignment or divergence of actors' objectives and the consequences attached to their actions, including rewards, penalties, performance measures, and other mechanisms that shape behavior. \emph{Alignment} here refers to objectives which may be closely aligned, partly divergent, or directly conflicting, reflecting the coalitional view of organizations, in which participants bargain over the goals of the firm (\hyperlink{ref:march1962}{March, 1962}). Multiple goals often coexist within organizations and are not necessarily fixed objectives (\hyperlink{ref:cyert1963-march}{Cyert \& March, 1963}; \hyperlink{ref:gaba2019-greve}{Gaba \& Greve, 2019}). Rather, these goals evolve as the relevant context changes both inside and outside the organization. \emph{Consequence-bearing mechanisms}, on the other hand, relate to the rewards, penalties, performance measures, and other behavior-inducing mechanisms that shape how actors pursue these objectives. Coordination between actors with aligned objectives differs significantly from coordination where cooperation must be induced, where actors withhold information or delay action, or where they pursue competing outcomes (\hyperlink{ref:jensen1976-meckling}{Jensen \& Meckling, 1976}; \hyperlink{ref:williamson1985}{Williamson, 1985}).

In AI-mediated organizational work, such structures may involve human actors as well as AI agents whose behavior is shaped by assigned objectives, evaluation criteria, or system-level constraints. The concept therefore does not presuppose human psychological motivation, but captures how differences in objectives and consequences condition organizational cooperation.

\emph{Environmental dynamism} concerns the stability of the conditions relevant to task performance. Environmental conditions often change while a task is underway, such as with technologies, customer demands, or decisions elsewhere in the organization. For example, a task can become significantly more organizationally demanding if the conditions on which the original plan was based suddenly change during task execution. Contingency and information-processing theories have long established the link between uncertainty and change in the environment and greater demands for information processing and adaptation (\hyperlink{ref:burton2004-obel}{Burton \& Obel, 2004}; \hyperlink{ref:lawrence1967-lorsch}{Lawrence \& Lorsch, 1967}). Environmental dynamism can also interact with the other dimensions in the taxonomy. For instance, in a highly dynamic environment, such as with a fast-growing startup, the effective control to make decisions can shift quickly, implying that this dynamism also has implications for authority structures. Furthermore, experienced actors in dynamic environments may be able to anticipate certain changes in task conditions, but in many instances dynamism entails change that is hard to predict (\hyperlink{ref:dess1984-beard}{Dess \& Beard, 1984}) or even impossible to predict.

The different dimensions outlined in the taxonomy describe distinct features of organizationally situated work, rather than points that can be scored on a single scale. For instance, involving more actors in a task does not necessarily imply greater organizational difficulty if their activities are largely independent. The configuration of dependencies, rather than their sheer number, thus shapes organizational difficulty.

\subsection*{Coordination requirements}
The second layer of the organizational embedding of work, coordination requirements, captures what the structure of work requires of the actors involved. Sequential interdependence calls for synchronization and management of hand-offs, while reciprocal interdependence requires continuous mutual adjustment between actors (\hyperlink{ref:thompson1967}{Thompson, 1967}; \hyperlink{ref:vandeven1976-delbecq}{Van de Ven et al., 1976}). Distributed information requires information acquisition and reconciliation, while distributed authority creates requirements for routing, authorization, and escalation. Temporal dependencies require sequencing relative to deadlines, waiting, and prioritization. Divergent interests can require negotiation and alignment, while changing environmental conditions call for anticipation and adaptation.

The mapping is not necessarily one-to-one, and multiple requirements may be present at once. For instance, distributed information combined with reciprocal interdependence may require both repeated information exchange and mutual adjustment. Ambiguous authority coupled with a tight deadline can make escalation especially important. Coordination requirements can thus arise from the combinations of structural features, as well as from the individual features themselves. For instance, \hyperlink{ref:vandeven1976-delbecq}{Van de Ven et al. (1976)} consider such features jointly, explicitly treating task uncertainty, interdependence, and unit size as determinants of alternative coordination modes rather than reducing coordination to a single mechanism.

\subsection*{Governance mechanisms}
The third layer consists of the different governance mechanisms that organizations use to meet these coordination requirements. Rules, routines, standards, plans, and predefined roles can program activities in advance. Information systems and communication channels allow actors to obtain and exchange knowledge. Authority and decision rights determine who may decide, approve, intervene, or resolve disagreement, while delegation reallocates responsibility for tasks and decisions. Direct interaction permits mutual adjustment as circumstances unfold. Monitoring, incentives, and escalation procedures address dependencies that cannot be managed through programming or information exchange alone (\hyperlink{ref:galbraith1974}{Galbraith, 1974}; \hyperlink{ref:march1958-simon}{March \& Simon, 1958}; \hyperlink{ref:mintzberg1979}{Mintzberg, 1979}; \hyperlink{ref:thompson1967}{Thompson, 1967}; \hyperlink{ref:williamson1985}{Williamson, 1985}).

Organization-design research treats these mechanisms as alternative and sometimes complementary responses to different demands arising from coordination. In settings where work is predictable, organizations can lean on rules and hierarchy, but as uncertainty rises, organizations can implement information systems and lateral relations to navigate higher demands for information processing (\hyperlink{ref:galbraith1974}{Galbraith, 1974}). Organizational effectiveness can thus be framed as the congruence between information-processing requirements and capacities (\hyperlink{ref:tushman1978-nadler}{Tushman \& Nadler, 1978}). For AI agents, the relevant issue is often not designing these mechanisms but recognizing and appropriately using those available in the organizational setting.

\subsection*{Performance outcomes}
Finally, the fourth layer concerns performance outcomes, capturing how the task is executed and what it achieves. Performance in organizationally situated work is multidimensional. \emph{Task outcomes} capture whether the output of the task execution satisfied the substantive requirements implied by the task. \emph{Coordination outcomes} capture whether relevant dependencies were handled successfully. \emph{Organizational conduct} concerns whether the agent respected decision rights, confidentiality, task ownership, escalation requirements, and other organizational constraints. \emph{Resource and time outcomes} capture whether the work was completed within the budget and time available.

A technically correct output does not necessarily imply successful performance within an organizational setting. An agent might produce the right strategic recommendation while bypassing required approval, disclosing information improperly, or acting too late for dependent activities to adjust. Furthermore, performance should be evaluated not only for the focal agent but also at the organizational level. The same agent might arrive at the right recommendation but execute the task in a way that imposes delays or generates unnecessary work for others, which can degrade overall organizational performance. An agent may thus perform adequately at the micro level without fulfilling the performance requirements of the organization. Organization-design research has long linked effectiveness to the fit between work requirements and coordination arrangements rather than to isolated task productivity (\hyperlink{ref:galbraith1974}{Galbraith, 1974}; \hyperlink{ref:tushman1978-nadler}{Tushman \& Nadler, 1978}).

\begin{table}[htbp]
\renewcommand{\thetable}{2}
\caption{The Taxonomy Dimensions of the Structure of Work}\label{tab:2}
\centering\small
\begin{tabularx}{\linewidth}{>{\raggedright\arraybackslash}p{0.22\linewidth}>{\raggedright\arraybackslash}X>{\raggedright\arraybackslash}p{0.3\linewidth}}
\toprule
\textbf{Dimension} & \textbf{Definition} & \textbf{References} \\
\midrule
\textbf{Task interdependence} & The interdependency between the focal task and other activities & \hyperlink{ref:thompson1967}{Thompson (1967)}; \hyperlink{ref:vandeven1976-delbecq}{Van de Ven et al. (1976)} \\
\addlinespace[0.6em]
\textbf{Information structure} & The distribution, accessibility, and character of task-relevant knowledge across organizational actors and information systems & \hyperlink{ref:hayek1945}{Hayek (1945)}; \hyperlink{ref:march1958-simon}{March \& Simon (1958)}; \hyperlink{ref:polanyi1966}{Polanyi (1966)}; \hyperlink{ref:galbraith1974}{Galbraith (1974)} \\
\addlinespace[0.6em]
\textbf{Authority structure} & The authority to request, decide, approve, override, and oblige & \hyperlink{ref:simon1947}{Simon (1947)}; \hyperlink{ref:jensen1992-meckling}{Jensen \& Meckling (1992)}; \hyperlink{ref:aghion1997-tirole}{Aghion \& Tirole (1997)} \\
\addlinespace[0.6em]
\textbf{Temporal structure} & The timing and order in which things must take place & \hyperlink{ref:thompson1967}{Thompson (1967)} \\
\addlinespace[0.6em]
\textbf{Incentive structure} & The alignment or divergence of actors' objectives and the consequences attached to their actions & \hyperlink{ref:march1962}{March (1962)}; \hyperlink{ref:cyert1963-march}{Cyert \& March (1963)}; \hyperlink{ref:jensen1976-meckling}{Jensen \& Meckling (1976)}; \hyperlink{ref:williamson1985}{Williamson (1985)} \\
\addlinespace[0.6em]
\textbf{Environmental dynamism} & The stability of the environmental conditions relevant to the task while work is underway & \hyperlink{ref:lawrence1967-lorsch}{Lawrence \& Lorsch (1967)}; \hyperlink{ref:burton2004-obel}{Burton \& Obel (2004)} \\
\bottomrule
\end{tabularx}
\end{table}

\subsection*{Scope of the taxonomy}
The developed taxonomy aims to provide a systematic classification of the organizational conditions that surround task performance. The six dimensions presented in the structure of work are intended to be analytically distinct, while still capturing the principal ideas identified in organization theory, including those around the division of labor and the integration of effort (\hyperlink{ref:lawrence1967-lorsch}{Lawrence \& Lorsch, 1967}; \hyperlink{ref:march1958-simon}{March \& Simon, 1958}; \hyperlink{ref:mintzberg1979}{Mintzberg, 1979}). They are also generally aligned with the universal problems of organizing (\hyperlink{ref:puranam2014-alexy}{Puranam et al., 2014}). While the dimensions are distinct, they can coexist and interact.

As in AI benchmarks, and given our emphasis on the task as the unit of analysis within an existing organizational context, we take the division and allocation of work as given. Human and non-human actors are assumed to already hold positions in the structure of work, and thus we do not focus on how they are recruited or assigned to tasks. Where such allocation takes place within the task itself, such as the delegation or hand-off of tasks, it is covered by authority structure and task interdependence.

Relatedly, the structure of work is not the only factor that shapes how a task is performed, as access to resources and tools, such as the right equipment, software, or capital, also matters. These resources and tools are treated as part of the technical task environment. They effectively determine whether an actor can perform the task at all, while the taxonomy describes the organizational conditions that shape the task's activities.

\FloatBarrier
\section*{ORGANIZATIONAL COVERAGE OF EXISTING AI BENCHMARKS}
If the former characterization of benchmarks as technically difficult---but organizationally silent---is right, it should be visible in benchmarks themselves. Moreover, the value of the taxonomy for AI benchmarking depends in part on whether it reveals differences that existing benchmark classifications obscure. As a result, we apply the taxonomy layers to existing AI benchmarks to examine which organizational conditions they represent and how deeply those conditions are incorporated into evaluation.

\subsection*{Benchmark sample and coding}
This section describes the process by which we collected the benchmarks and tagged them according to the dimensions of the taxonomy. We sampled benchmarks in three ways, attempting to capture what the field pays attention to: the benchmarks tracked by industry, recognized by the research community, and reported by frontier AI labs.

\begin{enumerate}
\item Industry indices: all benchmarks listed in two prominent public aggregator indices, the Artificial Analysis Intelligence Index and Epoch AI's Benchmarking Hub (\hyperlink{ref:artificialanalysis2026}{Artificial Analysis, 2026}; \hyperlink{ref:epochai2026}{Epoch AI, 2026}), as of October 8, 2026 (106 benchmarks).
\item Top-conference papers: benchmark papers at NeurIPS, ICML, ICLR, ACL, EMNLP and NAACL from 2023 to 2026 that were presented as an oral or spotlight, or received a paper award (321).
\item Lab model cards: benchmarks reported by at least two of five AI labs in their flagship model releases (166).
\end{enumerate}

Because a benchmark can fall into more than one of these groups, we tag each benchmark only once and count it only once. As a result, 593 group memberships yield 522 distinct benchmarks. We refer to these 522 benchmarks as the \emph{prominent benchmarks}. Because the three groups yield very similar aggregate results, we report them together (\hyperref[app:B]{Appendix B} breaks down the results by group). As a supplementary comparison discussed below, we also tag 523 conference benchmark papers classified as organizational or game-based.

We used a large language model (Claude Opus 5.5, at medium reasoning effort) to code each benchmark's paper against the taxonomy's definitions. For each of the six dimensions of the structure-of-work layer, we distinguish three levels of evidence in benchmark descriptions:

\begin{enumerate}
\item A benchmark may \emph{mention} a dimension anywhere in its paper, including in its motivation, related work or limitations.
\item It may \emph{claim} the dimension as part of the benchmark itself, in its tasks, its environment or its grader.
\item It may \emph{measure} performance on the dimension separately in its reported results, through a dedicated metric, an ablation that varies it, or a failure analysis that attributes errors to it.
\end{enumerate}

These distinctions matter because incorporation determines whether an organizational condition is present, while separate measurement determines whether performance with respect to that condition can be diagnosed per se. Every positive tag must be supported by a verbatim quote from the paper. We focus primarily on the six dimensions of work (Layer 1); for each dimension a benchmark claims or measures, we also tag the subsequent layers of the taxonomy, which we treat as more exploratory. \hyperref[app:A]{Appendix A} gives the full sampling and coding procedure, and \hyperref[app:E]{Appendix E} gives the exact definitions and prompts.

\subsection*{Mention, claim and measure}
\begin{figure}[htbp]
\renewcommand{\thefigure}{2}
\caption{Share of benchmarks that mention, claim and measure each dimension}\label{fig:2}
\centering
\includegraphics[width=0.95\linewidth]{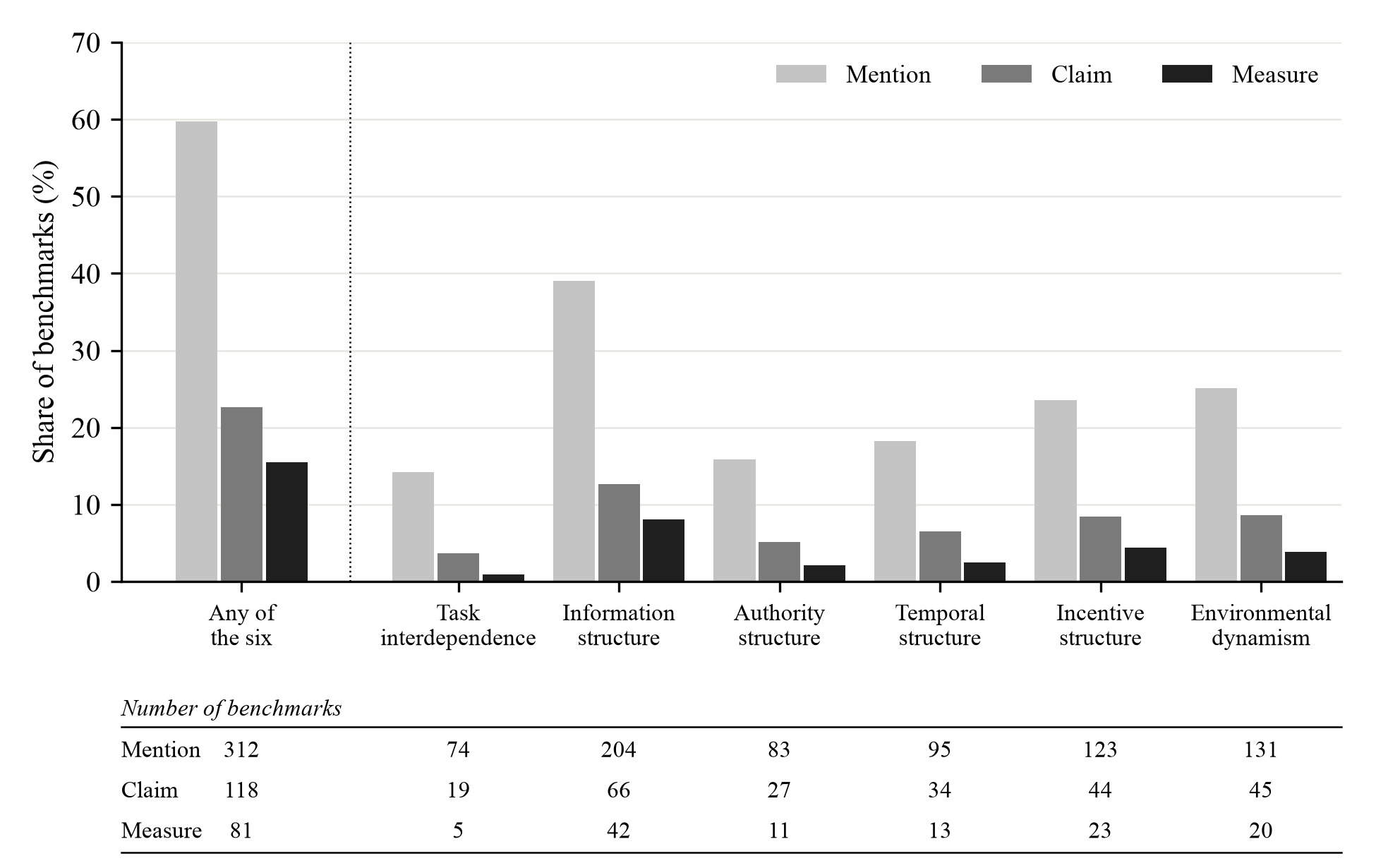}
\par\smallskip\begin{minipage}{\linewidth}\footnotesize\raggedright \emph{Note.} Benchmarks in the three prominent groups, each counted once (n = 522). Bars give the share of benchmarks; the counts are given beneath the figure. ``Any of the six'' counts benchmarks with the level on at least one dimension.\end{minipage}
\end{figure}

\hyperref[fig:2]{Figure 2} reports the share of benchmarks that mention, claim and measure each of the six dimensions. Only 23 percent of the benchmarks (118 of 522) claim any of the six dimensions, and about half of these (60) claim only one.

Two patterns emerge. Papers mention organizational conditions far more often than they claim to represent them in the benchmark, and represent them in the benchmark more often than they measure them. These conditions are therefore not obscure or unknown to benchmark authors; rather, they are not represented in the tasks. Additionally, coverage is highly concentrated: some dimensions, information structure in particular, are far more represented than others, such as task interdependence.

\begin{figure}[htbp]
\renewcommand{\thefigure}{3}
\caption{Share of benchmarks claiming at least one organizational dimension, by year of publication}\label{fig:3}
\centering
\includegraphics[width=0.95\linewidth]{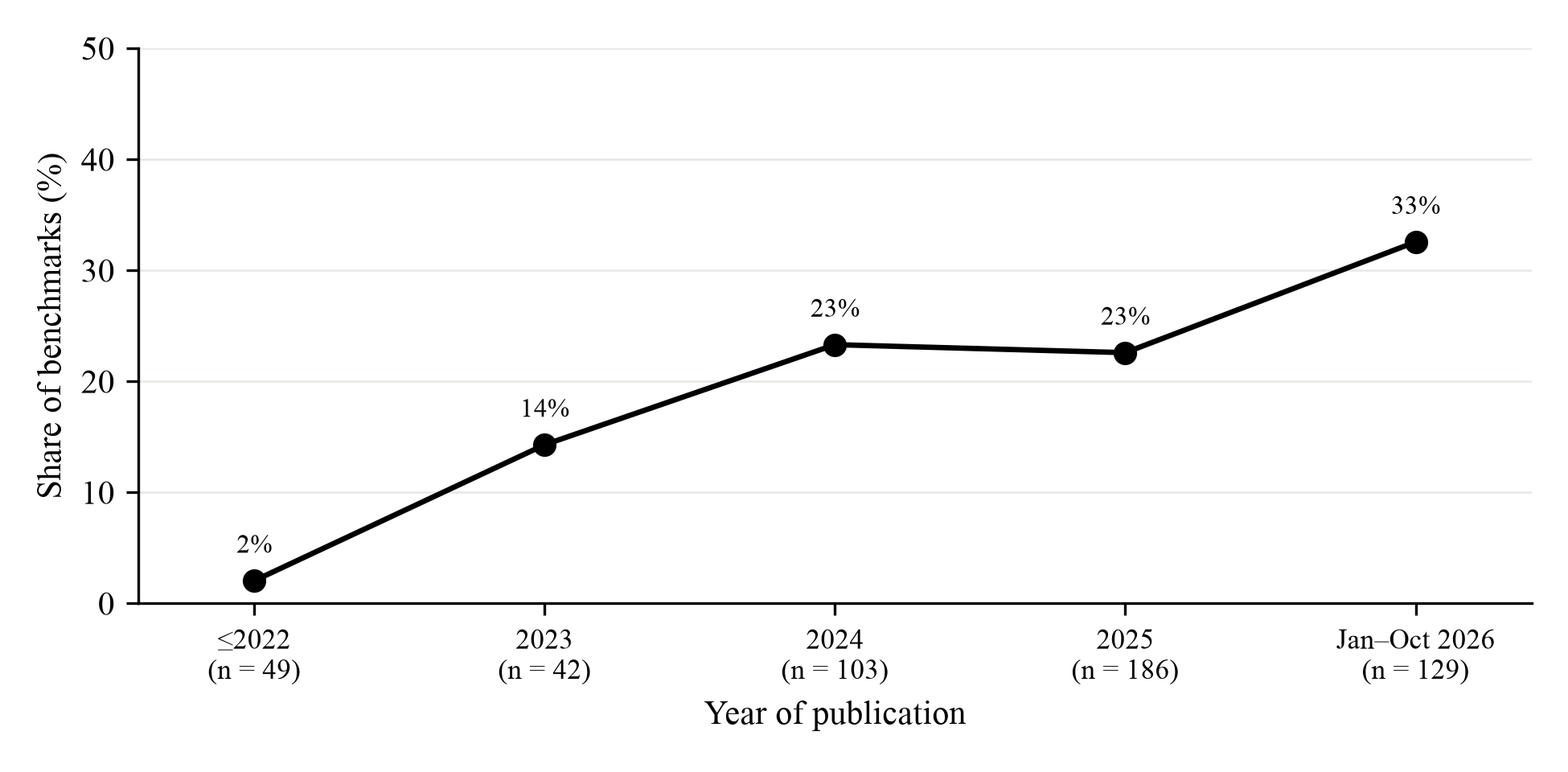}
\par\smallskip\begin{minipage}{\linewidth}\footnotesize\raggedright \emph{Note.} Benchmarks in the three prominent groups, each counted once. The number of benchmarks published in each year is shown under the year. 13 benchmarks without a dated source are excluded. The 2026 figure covers January to October.\end{minipage}
\end{figure}

Although still at a low level, organizational content is becoming more common in recent benchmarks (\hyperref[fig:3]{Figure 3}). Only 1 of the 49 benchmarks published before 2023 claims an organizational dimension, against 33 percent (42 of 129) of those published in 2026. Part of the rise reflects what the indices choose to track (rather than strictly what is being published), as the indices' recent versions emphasize agentic benchmarks.

\subsection*{Three example benchmarks}
\hyperref[tab:3]{Table 3} shows how three well-known benchmarks with organizational content were coded. $\tau^{2}$-bench (\hyperlink{ref:barres2025-dong}{Barres et al., 2025})\footnote{One of us (Victor Barres) is the first author of $\tau^{2}$-bench. We use it because it illustrates our point and we know it well; it was coded by the same automated procedure as every other benchmark.} is a customer-service benchmark that involves a simulated user as part of the environment. Our tagging flags the benchmark as measuring task interdependence separately, as the paper reports that moving from a setup where the agent acts alone to one where it must guide the user ``results in a substantial drop'' in success. TheAgentCompany (\hyperlink{ref:xu2024-song}{Xu et al., 2024}) tests an AI working as an employee in a simulated software company. It claims incentive structure, for example in a task where the agent ``negotiates with the simulated colleague to reduce the request'', and it also measures information structure. Vending-Bench 2 (\hyperlink{ref:andonlabs2025}{Andon Labs, 2025}) has an AI run a vending business and adds ``adversarial suppliers, delayed or failed deliveries, and customer refund demands'', which we code as environmental dynamism, but it places the agent in no organization: there are no colleagues, managers or approvals. \hyperref[tab:A1]{Table A1} (\hyperref[app:A]{Appendix A}) gives the quotations behind these codes. Reading cases such as these against the papers allowed us to iteratively refine the definitions during coding.

\begin{table}[htbp]
\renewcommand{\thetable}{3}
\caption{Highest level reached on each dimension, three example benchmarks}\label{tab:3}
\centering\small
\renewcommand{\arraystretch}{1.2}
\begin{tabularx}{\linewidth}{l>{\raggedright\arraybackslash}X>{\raggedright\arraybackslash}X>{\raggedright\arraybackslash}X}
\toprule
\textbf{Dimension} & \textbf{$\tau^{2}$-bench} & \textbf{TheAgentCompany} & \textbf{Vending-Bench 2} \\
\midrule
Task interdependence & Measure & Claim & None \\
Information structure & Measure & Measure & None \\
Authority structure & Claim & Mention & None \\
Temporal structure & Claim & Mention & Claim \\
Incentive structure & Mention & Claim & Claim \\
Environmental dynamism & Mention & Mention & Claim \\
\bottomrule
\end{tabularx}
\par\smallskip\begin{minipage}{\linewidth}\footnotesize\raggedright \emph{Note.} The levels are ordered: measure is higher than claim, and claim is higher than mention. All three benchmarks were coded automatically by the same procedure as every other benchmark; the codes describe what each paper states and are not a judgment of its quality.\end{minipage}
\end{table}

\subsection*{Subcomponents and situations}
\begin{figure}[htbp]
\renewcommand{\thefigure}{4}
\caption{Taxonomy subcomponents present among benchmarks claiming each dimension}\label{fig:4}
\centering
\includegraphics[width=0.82\linewidth]{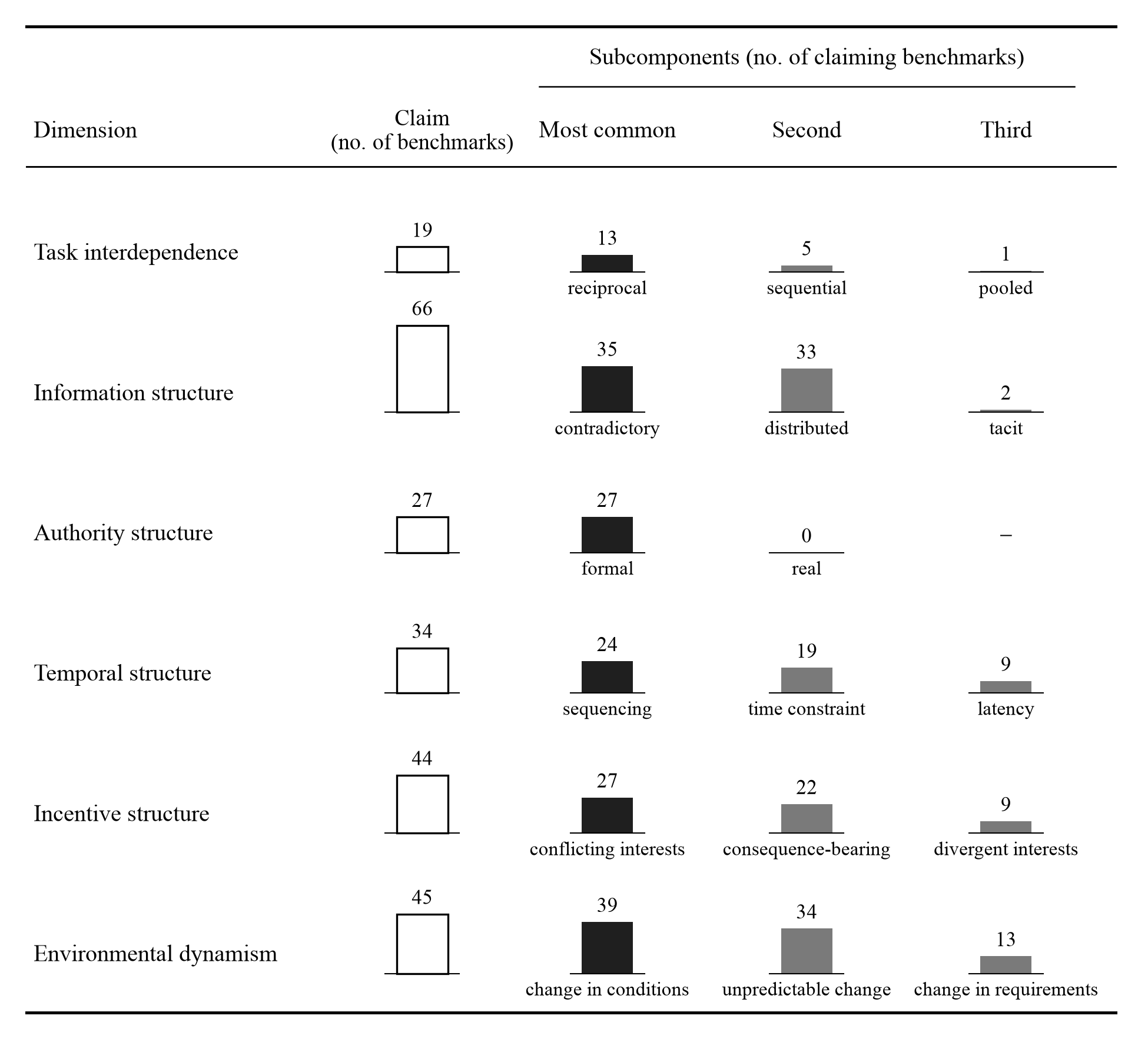}
\par\smallskip\begin{minipage}{\linewidth}\footnotesize\raggedright \emph{Note.} Benchmarks in the three prominent groups, each counted once. Each row is a dimension. The outlined bar gives the number of benchmarks claiming the dimension, and the filled bars give the number of those with the most common, second and third subcomponent; all bars share one scale. A benchmark can have more than one subcomponent, and baseline values (such as concentrated or explicit information) are not counted. Environmental dynamism also has change in resources (5) and change in roles (1).\end{minipage}
\end{figure}

The presence of a dimension by itself---e.g., information structure---does not say much about the variety of situations through which information structure is represented. For each dimension a benchmark claims, we therefore also record which of its taxonomy \emph{subcomponents} are present, and we find that most claims rest on a single form of the dimension (\hyperref[fig:4]{Figure 4}). Information structure, for example, rests almost entirely on contradictory and distributed information, while tacit knowledge appears in only two benchmarks.

Even within a subcomponent, benchmarks tend to represent the same situation the same way over and over. To get a sense of which scenarios recur, we sorted each claim into a rough list of common \emph{situations} for its dimension (\hyperref[app:C]{Appendix C}). Authority is the clearest case. All 27 benchmarks with authority structure rest on formal authority, and 11 of them on a written policy or set of rules the agent must follow, such as $\tau^{2}$-bench's (\hyperlink{ref:barres2025-dong}{Barres et al., 2025}) rule for when to transfer the user to a human agent (\hyperref[tab:A1]{Table A1}, \hyperref[app:A]{Appendix A}). The others mostly require confirmation from the person the agent serves, or a decision by a named role such as a manager or an investment committee (5 each).

Such situations do not reflect the variety of authority in organizational work. In a firm, authority is rarely one rule set out in advance. Instead, for example, there may be multiple overlapping authority chains, some of which are explicit and some of which are implicit. Additionally, who has authority may change dynamically, such as a purchase that crosses a spending limit and therefore requires manager approval. Apart from a simulated manager who posts purchasing limits and exceptions (OSWorld 2.0; \hyperlink{ref:yuan2026-zhou}{Yuan et al., 2026}) and an executive whose superiors cannot be reached (Agentic Misalignment; \hyperlink{ref:lynch2025-wright}{Lynch et al., 2025}), we find none of these situations in the prominent benchmarks.

The other dimensions show similar patterns, with each subcomponent usually expressed in one particular way. Information structure most often rests on contradictory sources the agent must reconcile within the material it is given (25 of 66), while knowledge held by colleagues appears in only 4. And of the 19 benchmarks that claim task interdependence, only two, TheAgentCompany and SWE-Lancer (\hyperlink{ref:miserendino2025-wang}{Miserendino et al., 2025}), involve colleagues in an organization; the rest involve a simulated user, a human collaborator or other agents in a game.

\subsection*{Subsequent layers}
For each dimension a benchmark claims or measures, we additionally tag the subsequent layers of the taxonomy: whether the task requires the corresponding coordination (Layer 2), whether the environment supplies a governance mechanism for it (Layer 3), and whether the grader checks the outcome (Layer 4). Each layer is coded against the part of the benchmark where it would appear: a coordination requirement is something the task requires of the agent, a governance mechanism is something the environment supplies, and a performance outcome is something the grader checks. For example, when a benchmark has information structure, we tag whether the agent must acquire and reconcile information (Layer 2) and whether the environment supplies information rules and standards (Layer 3). However, even the six Layer 1 dimensions require drawing together several literature bases, and the subsequent layers rest on finer distinctions that are necessarily less precise. The Layer 2--4 codes are doubly speculative, because these features are more granular in the benchmarks themselves than what a paper typically reports, and would thus be better tagged from the environment itself. This means that we likely miss features that a task contains but its paper does not describe, for example a grading check for approvals that the paper never mentions. We therefore treat these codes as exploratory (the items and prompts are given in \hyperref[app:E]{Appendix E}). Each layer is coded separately for every claimed dimension, so the figures in \hyperref[tab:4]{Table 4} are shares of the benchmarks claiming a given dimension, rather than a funnel in which a layer is only tagged if the preceding layers are present.

A missing tag, though, does not mean that a benchmark has no coordination requirement or no governance mechanism; it means only that none is formally specified. Where a benchmark specifies no governance mechanism, the agent falls back on its own judgment, much as an incomplete contract leaves the parties to fill the gaps with common sense. Papers almost never state that a mechanism is absent; they simply do not mention one. Of the 235 claimed dimensions, the paper says a governance mechanism is absent in only 8 cases and does not mention one in 143. A missing tag therefore means only that a mechanism is not made explicit. The taxonomy is meant to make such governance choices explicit, so that a benchmark either specifies a mechanism or deliberately leaves the matter to the agent.

Read this way, the subsequent layers are rarely made explicit (\hyperref[tab:4]{Table 4}). Papers state a governance mechanism or a grader check for only about a third of claimed dimensions, and incentives are the least developed. Where a governance mechanism is made explicit, it is usually a simple one, such as written rules or guidance, a chat channel for reaching colleagues, or a shared tracker: the prompt itself is treated as the governance. It also matters where the work is: coordination requirements and governance mechanisms arise mainly when work is divided among several actors, none of whom owns the whole process, whereas an agent sequencing its own steps faces a planning problem rather than a coordination problem. Overall, the subsequent layers are represented both sparsely and unevenly in AI benchmarks, with coordination requirements, governance mechanisms, and performance outcomes made explicit to very different degrees.

\begin{table}[htbp]
\renewcommand{\thetable}{4}
\caption{Subsequent layers stated by benchmarks that claim each dimension}\label{tab:4}
\centering\small
\renewcommand{\arraystretch}{1.2}
\begin{tabularx}{\linewidth}{l>{\centering\arraybackslash}X>{\centering\arraybackslash}X>{\centering\arraybackslash}X>{\centering\arraybackslash}X}
\toprule
\textbf{Dimension} & \textbf{Claim the dimension (Layer 1)} & \textbf{Task requires the coordination (Layer 2)} & \textbf{Environment supplies a mechanism (Layer 3)} & \textbf{Grader checks the outcome (Layer 4)} \\
\midrule
Task interdependence & 19 & 11 (58\%) & 17 (89\%) & 5 (26\%) \\
Information structure & 66 & 52 (79\%) & 23 (35\%) & 38 (58\%) \\
Authority structure & 27 & 9 (33\%) & 11 (41\%) & 15 (56\%) \\
Temporal structure & 34 & 22 (65\%) & 13 (38\%) & 16 (47\%) \\
Incentive structure & 44 & 7 (16\%) & 3 (7\%) & 4 (9\%) \\
Environmental dynamism & 45 & 29 (64\%) & 17 (38\%) & 10 (22\%) \\
\midrule
All claimed dimensions & 235 & 130 (55\%) & 84 (36\%) & 88 (37\%) \\
\bottomrule
\end{tabularx}
\par\smallskip\begin{minipage}{\linewidth}\footnotesize\raggedright \emph{Note.} Benchmarks in the three prominent groups, each counted once. The first column gives the number claiming each dimension; the other columns give the number (share of those claiming it) whose paper states each subsequent layer, with a supporting quotation found in the paper. The three subsequent-layer columns are separate shares of the same base, not successive filters. A missing tag means the paper does not state the layer: the benchmark may lack it, or it may have it without the paper describing it.\end{minipage}
\end{table}

\subsection*{Supplementary analysis: organizational and game-based benchmarks}
The three groups above describe what the field pays attention to, whether represented in industry indices, top-conference papers or lab model cards. As a supplementary analysis, we deliberately take the most generous selection rule we can: benchmarks chosen because they look organizational, either because they are explicitly organizational in nature or because they are game-based, since games share many organizational features, such as other actors with their own interests. We tag every benchmark paper at the six conferences above that is classified as organizational or game-based and whose full text is available, regardless of how it was presented (n = 523; 48 of these are also in the prominent sample, 42 of them among the top-conference papers). See \hyperref[app:A]{Appendix A} for a description of the tagging methodology. As expected, these benchmarks contain far more organizational content than the prominent benchmarks, on every dimension and at every level (\hyperref[tab:5]{Table 5}). Benchmarks with such content therefore do exist, but they are rarely among the most prominent. They also make up a growing share of benchmark research at the six conferences (\hyperref[fig:D1]{Figure D1}, \hyperref[app:D]{Appendix D}).

\begin{table}[htbp]
\renewcommand{\thetable}{5}
\caption{Benchmarks claiming each dimension: organizational and game-based benchmarks compared with prominent benchmarks}\label{tab:5}
\centering\small
\renewcommand{\arraystretch}{1.2}
\begin{tabularx}{\linewidth}{l>{\centering\arraybackslash}X>{\centering\arraybackslash}X>{\centering\arraybackslash}X}
\toprule
\textbf{Dimension} & \textbf{Prominent benchmarks (n = 522)} & \textbf{Organizational and game-based (n = 523)} & \textbf{Difference (percentage points)} \\
\midrule
Task interdependence & 19 (4\%) & 165 (32\%) & +28 \\
Information structure & 66 (13\%) & 272 (52\%) & +39 \\
Authority structure & 27 (5\%) & 86 (16\%) & +11 \\
Temporal structure & 34 (7\%) & 176 (34\%) & +27 \\
Incentive structure & 44 (8\%) & 194 (37\%) & +29 \\
Environmental dynamism & 45 (9\%) & 179 (34\%) & +26 \\
\midrule
Claims at least one & 118 (23\%) & 402 (77\%) & +54 \\
Mentions at least one & 312 (60\%) & 486 (93\%) & +33 \\
Measures at least one & 81 (16\%) & 300 (57\%) & +42 \\
\bottomrule
\end{tabularx}
\par\smallskip\begin{minipage}{\linewidth}\footnotesize\raggedright \emph{Note. Prominent benchmarks are those in the industry indices, top-conference papers and lab model cards.} Classification as organizational or game-based is from the paper's title and abstract (\hyperref[app:A]{Appendix A}). The difference is the share among organizational and game-based benchmarks minus the share among prominent benchmarks. The 48 benchmarks that are in both samples are counted in both columns; excluding them from both widens every difference, by at most 6 percentage points. \hyperref[app:D]{Appendix D} splits the organizational and game-based benchmarks by type.\end{minipage}
\end{table}

The two groups differ in the way one would expect. See \hyperref[app:D]{Appendix D} for more granular results. Game-based benchmarks are built with multi-agent difficulty in mind, so incentives and changing environments are about as common in game-based-only benchmarks as in organizational-only ones. However, games often do not have organizational structure as would exist in a firm, which is why hierarchy, decision rights and obligations, captured by authority structure, are rarely represented. The richest benchmarks are those classified as both, games set inside an organizational frame.

\begin{figure}[htbp]
\renewcommand{\thefigure}{C1}
\caption{Subcomponents present among organizational and game-based benchmarks claiming each dimension}\label{fig:C1}
\centering
\includegraphics[width=0.82\linewidth]{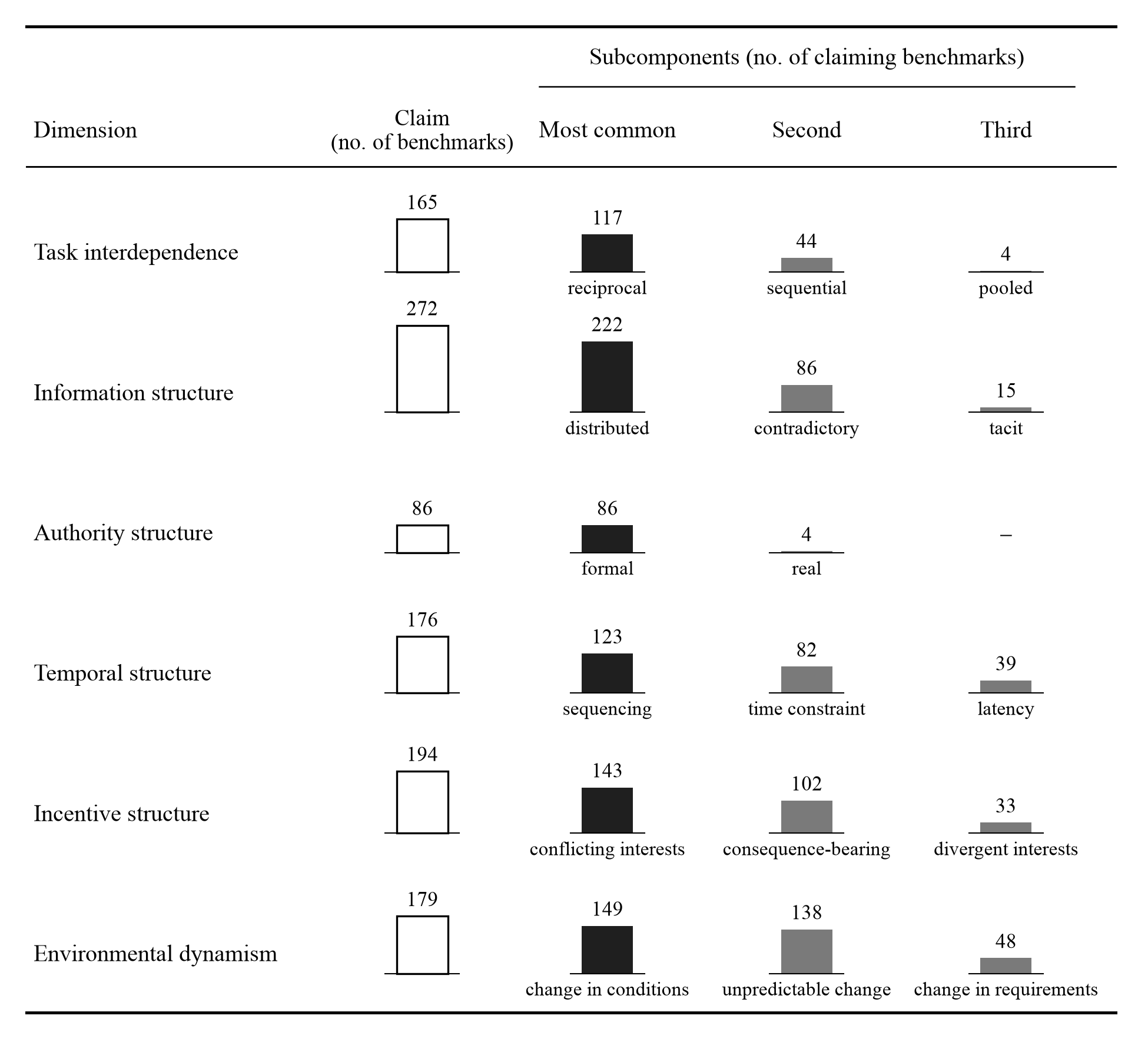}
\par\smallskip\begin{minipage}{\linewidth}\footnotesize\raggedright \emph{Note.} As \hyperref[fig:4]{Figure 4}, for the 523 organizational and game-based benchmarks. Environmental dynamism also has change in resources (27) and change in roles (7).\end{minipage}
\end{figure}

\begin{figure}[htbp]
\renewcommand{\thefigure}{D1}
\caption{Share of conference benchmark papers classified as organizational or game-based, by year}\label{fig:D1}
\centering
\includegraphics[width=0.95\linewidth]{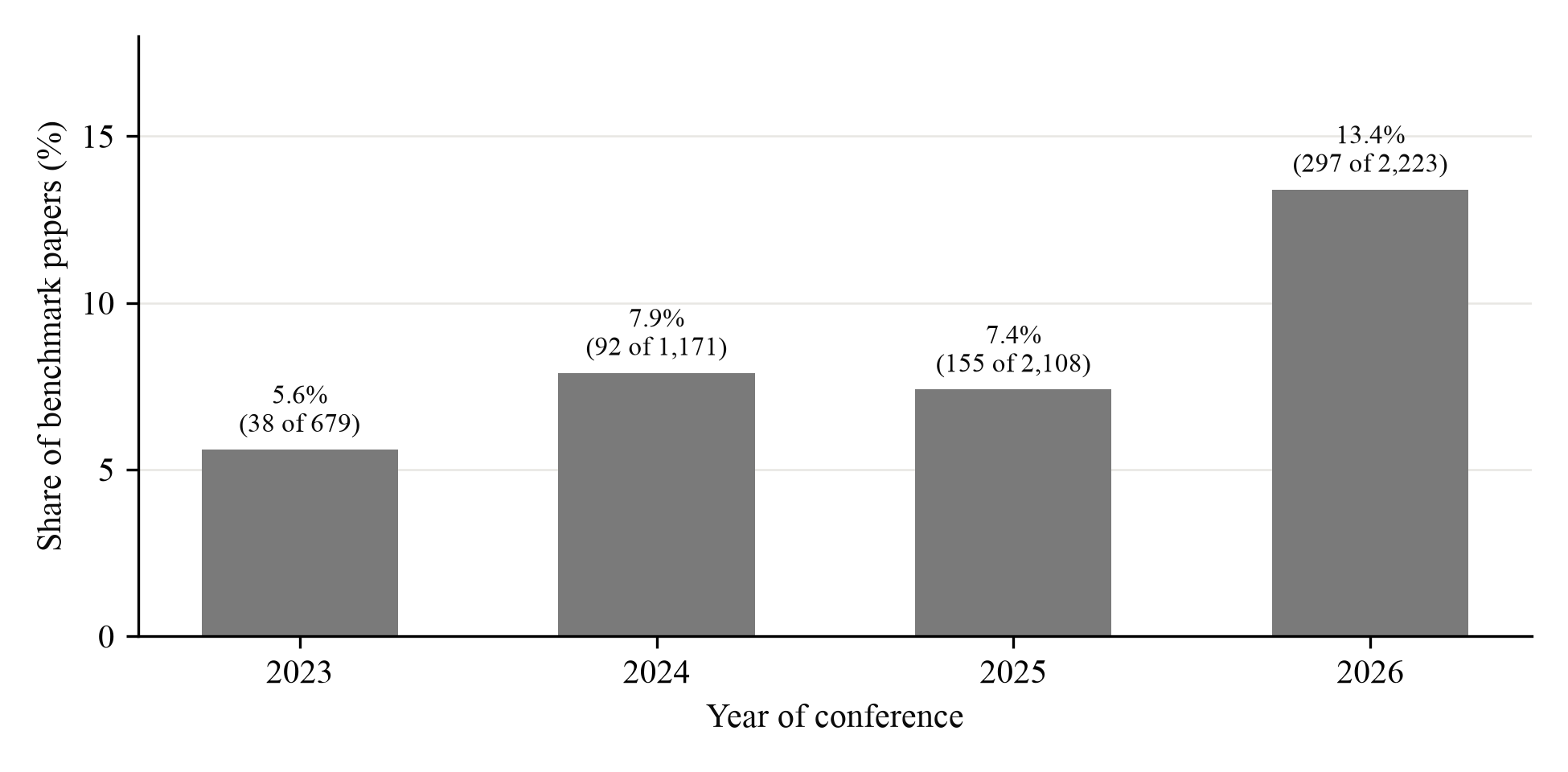}
\par\smallskip\begin{minipage}{\linewidth}\footnotesize\raggedright \emph{Note.} Share of benchmark papers at NeurIPS, ICML, ICLR, ACL, EMNLP and NAACL classified as organizational or game-based, by year. EMNLP 2026 had not been published at the time of collection, and NAACL was not held in 2023 or 2026.\end{minipage}
\end{figure}

Even within these two groups, though, we see the same concentration patterns, with some subcomponents and situations far more represented than others (\hyperref[fig:C1]{Figure C1}, \hyperref[app:C]{Appendix C}). All 86 benchmarks with authority structure represent formal authority, and real authority appears in only 4; and the knowledge the agent needs is held by colleagues or other members of an organization in only 19 of the 272 benchmarks with information structure. The subsequent layers are again rarely made explicit: papers state that the grader checks the corresponding outcome for only 21 to 49 percent of the benchmarks claiming each dimension. As in the prominent benchmarks, incentives are the least developed, with only about a quarter of papers (23 percent) stating that the agent must align or negotiate with others.

\FloatBarrier
\section*{DISCUSSION}

\subsection*{From task capability to organizational capability}
This paper begins from the simple separation between the difficulty of performing a focal task and the difficulty created by the organizational conditions surrounding its performance, which holds implications beyond AI benchmark design. Where AI models are adopted and embedded into existing, human-centered organizations, variation in effective performance may increasingly depend on whether agents can operate under conditions of interdependence, distributed knowledge, differentiated authority, temporal dependence, divergent interests, and environmental change---even as the technical capabilities of AI models increase. Technical capability and organizational capability should therefore be treated as distinct dimensions of AI performance rather than as points on a single capability scale. Progress on technical benchmarks thus need not imply progress along the organizational dimension.

Performance in organizational work, we argue, is tied to what we define as the \emph{organizational navigation capability:} an agent's ability to recognize how its actions are embedded in a changing structure of interdependence, distributed knowledge, authority, and commitments, and to adapt its behavior accordingly. The agent must hereby both \emph{recognize} its embeddedness within the organizational environment and dynamically \emph{adapt} actions in line with this understanding.

The capability sits with the agent, rather than with the conditions of the environment. The environment and agent are nonetheless linked: the environment sets forth the structures to navigate within, while the capability determines how well the agent manages. Furthermore, the structure of work conditions the organizational navigation capability, which therefore varies across different contexts and periods. The capability is not a fixed characteristic of the agent, but rather, it is relational and dynamic in nature, accounting for the changing structures of interdependence, knowledge, authority, and commitments that exist in organizations.

The capability does not appear to be a uniquely human feature. Where AI agents can exhibit understanding of the structure of work, they can adapt their action space accordingly. The organizational dimensions provide the axes of variability along which AI agents can fail, but also axes that they can adapt to.

Finally, organizational navigation capability should be distinguished from organizationally appropriate conduct. An agent may possess capabilities that allow it to acquire information, coordinate activities, and identify the dependencies surrounding a task, but exercise these capabilities in ways that violate decision rights, confidentiality requirements, or other organizational constraints. While organizational navigation capability refers to the agent's ability to recognize and respond to organizational conditions surrounding its work, \emph{organizational conduct} instead concerns whether the agent's actions are aligned with the organization's rules and norms governing such work. Both capability and conduct matter for an agent's organizational performance. The distinction between capability and conduct also becomes particularly relevant as agents acquire capabilities for navigating complex organizational environments without necessarily being reliable in how they exercise such capabilities.

\subsection*{What a benchmark score means}
Our results suggest that organizational conditions are largely missing from the benchmarks the field tracks most closely, and that even where they are represented, they are measured only about half the time. Across the prominent benchmarks we evaluate, fewer than one fifth separately measure an organizational dimension. While these organizational conditions may be implicitly embedded in a given benchmark and reflected in its aggregate score, such a score will not reveal whether a failure stems from the technical task or the organizational conditions surrounding it. Thus, representing an organizational condition in a benchmark and measuring an agent's ability to navigate it are different properties.

The results also point toward a field that has prioritized technical task difficulty in benchmarking. One way to see this is to place tasks on two axes, technical and organizational difficulty (\hyperref[fig:1]{Figure 1}). AI benchmarks have become harder---with many now posing problems that challenge even domain experts---but still tend to be well-specified, with the primary objectives, inputs, and authority to act often given up front. As a result, benchmarks such as GDPval (\hyperlink{ref:patwardhan2025-dias}{Patwardhan et al., 2025}) and the Remote Labor Index (\hyperlink{ref:mazeika2025-gatti}{Mazeika et al., 2025}) sit in the technically difficult quadrant: neither claims any of the six organizational dimensions, as in both benchmarks the agent works alone with the relevant materials supplied up front. This does not mean they are unrealistic. They are built around real professional work; it is simply not work that is organizationally difficult. Yet much other work within organizations is messier than this. Some of this work is technically simple but organizationally difficult, such as a nurse giving a routine dose of medication, which cannot be administered until a physician has ordered it and a pharmacist has checked it. Some is difficult on both counts, such as a lawyer drafting a complex contract while negotiating its terms with the client and opposing counsel.

This is not to say that these organizational conditions cannot be incorporated into AI benchmarks. As also surfaced by the results, some benchmarks are increasingly capturing organizational dimensions in their tasks. For instance, $\tau^{2}$-bench (\hyperlink{ref:barres2025-dong}{Barres et al., 2025}) isolates the cost of interdependence by comparing a setting where the agent acts alone with one where it must guide the user. The outlined taxonomy helps provide a conceptual architecture for incorporating organizational conditions into benchmarks more systematically. Because the taxonomy characterizes the organizational environment rather than the agent operating within it, an evaluation can hold the underlying technical task approximately constant while varying the organizational dimensions. This approach can help provide a basis for evaluating the organizational navigation capability of AI agents, capturing not only how well an agent can execute the task but also how well it performs as an embedded organizational actor.

\subsection*{What happens as AI gets better?}
Improvements in AI model capabilities on technical tasks in recent years have proceeded at a rapid pace, making future developments in capabilities appear likely. Recent releases of personal AI agents also show that agents are increasingly capable of not only supporting the user, but also executing tasks on their behalf, such as with booking flights or sending emails, which makes the ideas outlined in the taxonomy increasingly salient. Yet, as AI models improve on technical capabilities, and increasingly can be embedded as AI agents in organizations, it does not necessarily imply that organizational difficulty is eliminated. A swarm of capable agents with interdependent activities, differentiated knowledge, roles, authority, and incentives constitutes an organization in the sense captured by the taxonomy. In such an agentic organization, the organizational difficulty faced by agents is likely to resemble that faced by humans, such as determining who holds the relevant knowledge, who has the authority to override inputs, and who can sign off on strategic decisions. For instance, early evidence on multi-agent AI systems suggests that these systems also face problems of organizational design and coordination (\hyperlink{ref:cemri2025-pan}{Cemri et al., 2025}; see also \hyperlink{ref:arrieta2026-he}{Arrieta et al., 2026}). This does not necessarily imply that organizations will retain their current form, but rather that they will need to address the fundamental problems of organizing work.

At the same time, the computational capabilities of AI models are likely to enable faster execution of certain tasks, which holds implications for the structure of work. Where AI agents accelerate the production of task inputs and outputs, such as with code generation, drafting of documents, or automation of statistical analyses, the temporal structure of work may compress, reciprocal dependencies may need more frequent mutual adjustment, and the environmental conditions surrounding work may become more dynamic. Rather than doing away with organizational difficulty, the speed of task execution unlocked by AI agents may even increase it, or at least make it more salient for certain tasks.

\subsection*{Limitations and future research}
While we introduce the distinction between technical and organizational task difficulty, and through our results document that AI benchmarks primarily focus on the former, we also acknowledge the rapid progress made in both AI models and benchmarks. AI models and agents will continue to evolve in the years ahead, which could influence the validity of the ideas and taxonomy outlined in this paper. At the same time, the outlined ideas are derived from organization theory developed through the decades and could stand the test of time---even in the age of AI.

The outlined taxonomy is based on theory that was developed for human organizations, which represents a limitation. One could hypothesize that agentic AI organizations differ from human organizations in notable ways, in part because some of the foundational assumptions of organizations---bounded rationality, knowledge specialization, hierarchy, and self-interest (\hyperlink{ref:garicano2000}{Garicano, 2000}; \hyperlink{ref:hayek1945}{Hayek, 1945}; \hyperlink{ref:simon1955}{Simon, 1955}; \hyperlink{ref:williamson1985}{Williamson, 1985})---might be altered with an agentic workforce. Recent events have shown the notable ability of AI agents to coordinate at significant scale, such as with the OpenAI--Hugging Face hacking incident in 2026 (\hyperlink{ref:metr2026}{METR, 2026}). Our ideas therefore anchor on human-centered organizations, where AI agents enter alongside existing human employees, and this focus is also a potential limitation of the paper.

Finally, the tagging methodology focuses mainly on the first layer of the organizational embedding of work, the structure of work. The results for the other layers are more tentative, since these layers are tagged only for the dimensions a benchmark claims and only for whether the paper makes them explicit, not whether the benchmark contains them. The layers are, however, to a certain extent interconnected, and some, but not necessarily all, can be present in a given AI benchmark.

The ideas outlined in this paper provide a springboard from which the taxonomy can be further developed. Future research could provide a further breakdown of the organizational subcomponents that underpin the six dimensions, as well as further elaborate Layers 2--4 in the context of AI models. Furthermore, research could explore whether the ideas outlined herein also apply to agentic forms of organizing. The taxonomy also provides a call to action to the field of AI benchmarking: technical capabilities are one side of the equation, while organizational capabilities represent the other, less explored side. This calls for AI benchmarks that increasingly focus on the organizational conditions, and for research on how AI models themselves can acquire the necessary organizational navigation capability to navigate these. Finally, as AI agents become more capable and increasingly integrated into organizational work, more research is warranted on the division of labor between humans and AI.

\FloatBarrier
\section*{CONCLUSION}
The central implication of this paper's analysis is that the frontier of AI capability cannot be characterized by technical task difficulty alone. The economically valuable work that goes on within organizations is also characterized by organizational task difficulty, which is an insufficiently recognized dimension of AI capability. As AI systems continue to improve and move from answering questions to performing work for humans, AI benchmarks must also account for the organizational conditions in which tasks are embedded. By drawing on organization theory, we distinguish four layers of the organizational embedding of work, namely the structure of work, coordination requirements, governance mechanisms, and performance outcomes. Furthermore, by developing a taxonomy of the structure of work and applying it to existing AI benchmarks, we find limited and uneven coverage of organizational conditions incorporated into such benchmarks. The taxonomy and analysis of this paper are intended as a first step toward making organizational work measurable for the evaluation of frontier AI systems.

\FloatBarrier

\FloatBarrier
\clearpage
\section*{APPENDIX}

\FloatBarrier
\subsection*{Appendix A. Benchmark sample and coding}\phantomsection\label{app:A}
This appendix describes the process by which we collected the benchmarks, processed them, and tagged them according to the dimensions of the organizational taxonomy.

\subsubsection*{The collection}
We sampled 997 benchmarks in total, organized into four groups. Importantly, this sampling was not meant to be exhaustive or definitive, but rather a fair attempt at finding both prominent benchmarks and benchmarks that plausibly would have organizational content. The first three groups capture what the field pays attention to: the benchmarks tracked by industry, recognized by the research community, and reported by frontier AI labs. The fourth captures benchmarks that plausibly have organizational content, which we analyze as a supplementary comparison. Each group is defined by its own criterion, and a benchmark that meets more than one criterion is included in each of those groups; for example, a benchmark that appears in an industry index and is reported by several labs counts in both. We tag each benchmark once. In the main text we combine the three prominent groups and count each benchmark once (522 distinct benchmarks from 593 group memberships); \hyperref[app:B]{Appendix B} reports the groups separately.

\begin{enumerate}
\item Industry indices. We assembled all benchmarks listed in two prominent public aggregator indices, the Artificial Analysis Intelligence Index and Epoch AI's Benchmarking Hub (\hyperlink{ref:artificialanalysis2026}{Artificial Analysis, 2026}; \hyperlink{ref:epochai2026}{Epoch AI, 2026}), as of October 8, 2026, attempting to gather the benchmarks the industry tracks most closely (106 benchmarks).
\item Top-conference papers. We collected all papers accepted at NeurIPS (including the Datasets and Benchmarks track), ICML and ICLR, and at ACL, EMNLP and NAACL (including Findings), from 2023 to 2026, approximately 66,000 papers. We used a language model to classify every paper from its title and abstract, with no keyword pre-screen, keeping those whose primary contribution is a new benchmark, evaluation dataset or evaluation environment (6,181 papers; prompt in \hyperref[app:E]{Appendix E}). Of these, we retained the papers presented as an oral or spotlight at NeurIPS, ICML or ICLR, or that received a paper award at an ACL venue, as a signal of recognition by the research community. We tagged the 321 for which the full text was available.
\item Lab model cards. We collected the launch posts, system cards, model cards and technical reports for the flagship model releases of OpenAI, Anthropic, Google, xAI and Meta from January 2023 to October 2026 (267 documents), and recorded every benchmark they report, treating versions and subsets of the same benchmark as one. We retained benchmarks reported by at least two of the five labs, excluding human exams (such as the GRE), leaderboards, and internal evaluations with no public description (166 benchmarks).
\item Organizational and game-based benchmarks. We also assembled benchmarks that were plausibly more relevant to organizational work. Using the same classification of top-conference papers described above, we asked the model, for each benchmark paper, whether it is organizational (``the benchmark explicitly places the agent in a workplace or organizational setting, or with other actors whose roles, decisions, information or interests matter'') or game-based (``the benchmark is built on games or game-like play''), judging from its title and abstract (prompt in \hyperref[app:E]{Appendix E}). We retained every benchmark paper classified as organizational, game-based or both, regardless of how it was presented at the conference: 582 papers (434 organizational and 253 game-based, of which 105 are both). We tagged the 523 for which the full text was available. These labels only select the candidates: whether a benchmark actually contains an organizational dimension is then determined by the full-paper tagging.
\end{enumerate}

\subsubsection*{Tagging and processing}
We tagged each benchmark along the taxonomy. We used the benchmark's paper as the core source document. We did not use code, task files, or leaderboards, as many benchmarks do not release these materials publicly, whereas a benchmark's paper is a near-universal source from which to analyze it. We matched each benchmark to its paper using its arXiv identifier, found through the conference record, Semantic Scholar or a title search, and checked every paper's title against the benchmark; for 58 conference papers not on arXiv, we used the published version from the ACL Anthology or the conference proceedings. There were 39 benchmarks for which we could not find a paper in these sources. For these, we used the benchmark's accompanying web page, blog post, technical report or repository.

\subsubsection*{Coding procedure}
We used a large language model (Claude Opus 5.5, at medium reasoning effort) to classify and code the papers, in order to enable labeling at scale. The labeling process was the same for all benchmarks included. The model saw only the paper (or, where none exists, the benchmark's web page, blog post, technical report or repository) and the taxonomy's definitions. We developed these definitions iteratively, reviewing the tagging outputs against the concepts and refining the definitions where the tags did not capture what we intended. See \hyperref[app:E]{Appendix E} for the exact definitions we used in the tagging.

\subsubsection*{What we code}
For each of the six dimensions of the structure-of-work layer, we distinguish three levels of evidence in benchmark descriptions. A benchmark may (1) mention a dimension anywhere in its paper, including in its motivation, related work or limitations; (2) claim the dimension as part of the benchmark itself, in its tasks, its environment or its grader; or (3) measure performance on the dimension separately in its reported results, through a dedicated metric, an ablation that varies it, or a failure analysis that attributes errors to it. Note that a claim counts only if the dimension is established under our definitions; for example, an agent reading many supplied documents does not by itself establish information structure. For each dimension a benchmark claims, we also record which of its subcomponents are present, for example whether authority is formal or real, or whether interdependence is pooled, sequential or reciprocal. We do not tag the rate at which the environment changes, which the definitions leave qualitative.

These categories distinguish recognition of an organizational condition from its reported incorporation in the benchmark and, more importantly, from the ability to identify whether an agent handles that condition successfully. This distinction is important because an organizational condition may affect task performance without being separately measured. A benchmark may, for example, require an agent to obtain approval from another actor, thereby incorporating authority structure. If failure to obtain approval is folded into an aggregate task-success score, however, the benchmark cannot establish whether poor performance reflects a failure to navigate authority rather than some other aspect of the task. Hence, incorporation determines whether an organizational condition is present, while separate measurement determines whether performance with respect to that condition can be diagnosed. Every positive code must be supported by a verbatim quote from the paper, which we check against the paper text.

For each dimension a benchmark claims or measures, we also code the subsequent layers of the taxonomy. We ask whether the benchmark's tasks require the corresponding coordination (Layer 2), whether its environment supplies a governance mechanism for it (Layer 3), and whether its grader checks the resulting outcome (Layer 4). These codes record whether the paper makes each layer explicit, not whether the benchmark lacks it. For information structure, for example, we ask whether the agent must acquire and reconcile information held by others, whether the environment provides information rules or systems for doing so, and whether the grader checks that correct and complete information was obtained. Because the subsequent layers are developed in less detail than the structure of work, these codes are more tentative and are intended to indicate direction rather than to support precise comparisons. The full procedure, definitions and questions are given in \hyperref[app:E]{Appendix E}.

\subsubsection*{Situations}
To describe how a benchmark realizes each claimed dimension, we classified each claim by its supporting quotation into one of a short list of situations for that dimension (for example, ``written policy or rules the agent must follow'' or ``approval or decision by a named role'' for authority), with ``other'' where none fits. \hyperref[app:C]{Appendix C} gives the full lists and counts.

\begin{table}[htbp]
\renewcommand{\thetable}{A1}
\caption{Example quotations supporting the codes in Table 3}\label{tab:A1}
\centering\small
\begin{tabularx}{\linewidth}{>{\raggedright\arraybackslash}p{0.19\linewidth}>{\raggedright\arraybackslash}p{0.16\linewidth}>{\raggedright\arraybackslash}p{0.09\linewidth}>{\raggedright\arraybackslash}X}
\toprule
\textbf{Benchmark} & \textbf{Dimension} & \textbf{Level} & \textbf{Supporting paper quotation} \\
\midrule
$\tau^{2}$-bench & Task interdependence & Measure & ``shifting from no user operation (No-User) to a collaborative setup (Default) where the agent must guide the user results in a substantial drop in pass\textasciicircum{}1 (18\% drop for gpt-4.1 and 25\% drop for o4-mini).'' \\
\addlinespace[0.6em]
$\tau^{2}$-bench & Authority structure & Claim & ``You should transfer the user to a human agent if and only if the request cannot be handled within the scope of your actions.'' \\
\addlinespace[0.6em]
TheAgentCompany & Information structure & Measure & ``RocketChat is where all social interaction with peers occurs, and the low scores suggest that LLMs still lack communication skills.'' \\
\addlinespace[0.6em]
TheAgentCompany & Incentive structure & Claim & ``After calculating that the requested items exceed the budget, the agent negotiates with the simulated colleague to reduce the request'' \\
\addlinespace[0.6em]
Vending-Bench 2 & Environmental dynamism & Claim & ``version 2 adds real-world messiness such as adversarial suppliers, delayed or failed deliveries, and customer refund demands'' \\
\bottomrule
\end{tabularx}
\end{table}

\FloatBarrier
\subsection*{Appendix B. Results by group}\phantomsection\label{app:B}
The three prominent groups give very similar results (\hyperref[tab:B1]{Table B1}). The overlap between groups does not drive this: benchmarks in only one group claim a dimension about as often as those shared with another (industry indices 26 against 20 percent, lab model cards 24 against 18 percent; only 13 top-conference papers are shared). In all, the 593 group memberships are 522 distinct benchmarks: 454 are in one group, 65 in two and 3 in all three.

\begin{table}[htbp]
\renewcommand{\thetable}{B1}
\caption{Benchmarks mentioning, claiming and measuring each dimension, by group}\label{tab:B1}
\centering\small
\renewcommand{\arraystretch}{1.2}
\begin{tabularx}{\linewidth}{l>{\centering\arraybackslash}X>{\centering\arraybackslash}X>{\centering\arraybackslash}X}
\toprule
\textbf{Dimension} & \textbf{Industry indices (n = 106)} & \textbf{Top-conference papers (n = 321)} & \textbf{Lab model cards (n = 166)} \\
\midrule
Task interdependence & 3 (3\%) & 14 (4\%) & 4 (2\%) \\
Information structure & 13 (12\%) & 40 (12\%) & 23 (14\%) \\
Authority structure & 7 (7\%) & 12 (4\%) & 13 (8\%) \\
Temporal structure & 10 (9\%) & 19 (6\%) & 12 (7\%) \\
Incentive structure & 8 (8\%) & 26 (8\%) & 15 (9\%) \\
Environmental dynamism & 8 (8\%) & 32 (10\%) & 11 (7\%) \\
\midrule
Claims at least one & 24 (23\%) & 72 (22\%) & 36 (22\%) \\
Mentions at least one & 70 (66\%) & 188 (59\%) & 103 (62\%) \\
Measures at least one & 14 (13\%) & 53 (17\%) & 21 (13\%) \\
\bottomrule
\end{tabularx}
\par\smallskip\begin{minipage}{\linewidth}\footnotesize\raggedright \emph{Note.} Each group is counted in full, so a benchmark in two groups appears in both columns. Dimension rows give benchmarks claiming the dimension.\end{minipage}
\end{table}

\FloatBarrier
\subsection*{Appendix C. Subcomponents and situations in full}\phantomsection\label{app:C}
\begin{table}[htbp]
\renewcommand{\thetable}{C1}
\caption{Situations through which benchmarks realize each claimed dimension}\label{tab:C1}
\centering\small
\begin{tabularx}{\linewidth}{>{\raggedright\arraybackslash}p{0.17\linewidth}>{\raggedright\arraybackslash}X>{\raggedright\arraybackslash}X}
\toprule
\textbf{Dimension} & \textbf{Prominent benchmarks (n = 522)} & \textbf{Organizational and game-based benchmarks (n = 523)} \\
\midrule
Task interdependence & Claiming: 19 \par \hangindent=0.8em\hangafter=1 \textbullet{} human collaborator (6) \par \hangindent=0.8em\hangafter=1 \textbullet{} simulated user (6) \par \hangindent=0.8em\hangafter=1 \textbullet{} other agents or players in a game (4) \par \hangindent=0.8em\hangafter=1 \textbullet{} colleagues (2) \par \hangindent=0.8em\hangafter=1 \textbullet{} other (1) & Claiming: 165 \par \hangindent=0.8em\hangafter=1 \textbullet{} other agents or players in a game (108) \par \hangindent=0.8em\hangafter=1 \textbullet{} simulated user (26) \par \hangindent=0.8em\hangafter=1 \textbullet{} human collaborator (20) \par \hangindent=0.8em\hangafter=1 \textbullet{} colleagues (7) \par \hangindent=0.8em\hangafter=1 \textbullet{} other (4) \\
\addlinespace[0.6em]
Information structure & Claiming: 66 \par \hangindent=0.8em\hangafter=1 \textbullet{} conflicting sources (25) \par \hangindent=0.8em\hangafter=1 \textbullet{} simulated user who must be asked (16) \par \hangindent=0.8em\hangafter=1 \textbullet{} other players in a game (8) \par \hangindent=0.8em\hangafter=1 \textbullet{} files or web pages (5) \par \hangindent=0.8em\hangafter=1 \textbullet{} colleagues (4) \par \hangindent=0.8em\hangafter=1 \textbullet{} database or system (3) \par \hangindent=0.8em\hangafter=1 \textbullet{} other (5) & Claiming: 272 \par \hangindent=0.8em\hangafter=1 \textbullet{} simulated user who must be asked (94) \par \hangindent=0.8em\hangafter=1 \textbullet{} other players in a game (72) \par \hangindent=0.8em\hangafter=1 \textbullet{} conflicting sources (55) \par \hangindent=0.8em\hangafter=1 \textbullet{} colleagues (19) \par \hangindent=0.8em\hangafter=1 \textbullet{} files or web pages (6) \par \hangindent=0.8em\hangafter=1 \textbullet{} database or system (6) \par \hangindent=0.8em\hangafter=1 \textbullet{} human collaborator (1) \par \hangindent=0.8em\hangafter=1 \textbullet{} other (19) \\
\addlinespace[0.6em]
Authority structure & Claiming: 27 \par \hangindent=0.8em\hangafter=1 \textbullet{} written policy or rules (11) \par \hangindent=0.8em\hangafter=1 \textbullet{} confirmation from the person served (5) \par \hangindent=0.8em\hangafter=1 \textbullet{} named role (5) \par \hangindent=0.8em\hangafter=1 \textbullet{} hierarchy among agents or players (2) \par \hangindent=0.8em\hangafter=1 \textbullet{} other (4) & Claiming: 86 \par \hangindent=0.8em\hangafter=1 \textbullet{} written policy or rules (25) \par \hangindent=0.8em\hangafter=1 \textbullet{} hierarchy among agents or players (22) \par \hangindent=0.8em\hangafter=1 \textbullet{} named role (16) \par \hangindent=0.8em\hangafter=1 \textbullet{} confirmation from the person served (13) \par \hangindent=0.8em\hangafter=1 \textbullet{} other (10) \\
\addlinespace[0.6em]
Temporal structure & Claiming: 34 \par \hangindent=0.8em\hangafter=1 \textbullet{} required order set by a process (13) \par \hangindent=0.8em\hangafter=1 \textbullet{} deadline set in the task (8) \par \hangindent=0.8em\hangafter=1 \textbullet{} turn or round limits (7) \par \hangindent=0.8em\hangafter=1 \textbullet{} waiting for another actor (3) \par \hangindent=0.8em\hangafter=1 \textbullet{} interruptions (2) \par \hangindent=0.8em\hangafter=1 \textbullet{} other (1) & Claiming: 176 \par \hangindent=0.8em\hangafter=1 \textbullet{} turn or round limits (68) \par \hangindent=0.8em\hangafter=1 \textbullet{} required order set by a process (58) \par \hangindent=0.8em\hangafter=1 \textbullet{} deadline set in the task (23) \par \hangindent=0.8em\hangafter=1 \textbullet{} waiting for another actor (16) \par \hangindent=0.8em\hangafter=1 \textbullet{} interruptions (3) \par \hangindent=0.8em\hangafter=1 \textbullet{} other (8) \\
\addlinespace[0.6em]
Incentive structure & Claiming: 44 \par \hangindent=0.8em\hangafter=1 \textbullet{} adversary or attacker (12) \par \hangindent=0.8em\hangafter=1 \textbullet{} budget or resources (9) \par \hangindent=0.8em\hangafter=1 \textbullet{} user who wants what the agent should not give (6) \par \hangindent=0.8em\hangafter=1 \textbullet{} points in a game (4) \par \hangindent=0.8em\hangafter=1 \textbullet{} opponent in a game (3) \par \hangindent=0.8em\hangafter=1 \textbullet{} negotiation counterpart (3) \par \hangindent=0.8em\hangafter=1 \textbullet{} other (7) & Claiming: 194 \par \hangindent=0.8em\hangafter=1 \textbullet{} opponent in a game (48) \par \hangindent=0.8em\hangafter=1 \textbullet{} points in a game (41) \par \hangindent=0.8em\hangafter=1 \textbullet{} adversary or attacker (25) \par \hangindent=0.8em\hangafter=1 \textbullet{} negotiation counterpart (23) \par \hangindent=0.8em\hangafter=1 \textbullet{} budget or resources (22) \par \hangindent=0.8em\hangafter=1 \textbullet{} user who wants what the agent should not give (13) \par \hangindent=0.8em\hangafter=1 \textbullet{} other (22) \\
\addlinespace[0.6em]
Environmental dynamism & Claiming: 45 \par \hangindent=0.8em\hangafter=1 \textbullet{} world or system changes on its own (27) \par \hangindent=0.8em\hangafter=1 \textbullet{} new requirements arrive mid-task (8) \par \hangindent=0.8em\hangafter=1 \textbullet{} other players' moves (6) \par \hangindent=0.8em\hangafter=1 \textbullet{} information injected mid-task (4) & Claiming: 179 \par \hangindent=0.8em\hangafter=1 \textbullet{} world or system changes on its own (83) \par \hangindent=0.8em\hangafter=1 \textbullet{} other players' moves (49) \par \hangindent=0.8em\hangafter=1 \textbullet{} new requirements arrive mid-task (37) \par \hangindent=0.8em\hangafter=1 \textbullet{} information injected mid-task (6) \par \hangindent=0.8em\hangafter=1 \textbullet{} other (4) \\
\bottomrule
\end{tabularx}
\end{table}

\FloatBarrier
\subsection*{Appendix D. Organizational and game-based benchmarks by type}\phantomsection\label{app:D}
\begin{table}[htbp]
\renewcommand{\thetable}{D1}
\caption{Organizational and game-based benchmarks mentioning, claiming and measuring each dimension, by type}\label{tab:D1}
\centering\small
\renewcommand{\arraystretch}{1.2}
\begin{tabularx}{\linewidth}{l>{\centering\arraybackslash}X>{\centering\arraybackslash}X>{\centering\arraybackslash}X}
\toprule
\textbf{Dimension} & \textbf{Organizational only (n = 294)} & \textbf{Both (n = 94)} & \textbf{Game-based only (n = 135)} \\
\midrule
Task interdependence & 79 (27\%) & 70 (74\%) & 16 (12\%) \\
Information structure & 171 (58\%) & 70 (74\%) & 31 (23\%) \\
Authority structure & 62 (21\%) & 22 (23\%) & 2 (1\%) \\
Temporal structure & 85 (29\%) & 64 (68\%) & 27 (20\%) \\
Incentive structure & 82 (28\%) & 72 (77\%) & 40 (30\%) \\
Environmental dynamism & 84 (29\%) & 53 (56\%) & 42 (31\%) \\
\midrule
Claims at least one & 241 (82\%) & 90 (96\%) & 71 (53\%) \\
Mentions at least one & 289 (98\%) & 94 (100\%) & 103 (76\%) \\
Measures at least one & 193 (66\%) & 70 (74\%) & 37 (27\%) \\
\bottomrule
\end{tabularx}
\par\smallskip\begin{minipage}{\linewidth}\footnotesize\raggedright \emph{Note.} Classification is from the paper's title and abstract (\hyperref[app:A]{Appendix A}); ``Both'' are benchmarks classified as both organizational and game-based. Dimension rows give benchmarks claiming the dimension.\end{minipage}
\end{table}

\FloatBarrier
\subsection*{Appendix E. Tagging procedure, definitions and prompts}\phantomsection\label{app:E}
All tagging was done by a large language model (Claude Opus 5.5, medium reasoning effort). Definitions and questions below are the exact text the model received. Layer 1 definitions and notes are reproduced verbatim. Layers 2--4 item names and descriptions are verbatim from the four-layer taxonomy.

\subsubsection*{E1. Coding procedure}\phantomsection\label{sec:E1}
\begin{enumerate}
\item \textbf{Initial pass (Layer 1).} We feed the full paper text into the model's context, together with the definitions of the six dimensions, their subcomponents and the tagging notes in \hyperref[sec:E2]{section E2}. For each dimension, the model returns a yes/no answer to the following questions:
\begin{itemize}
\item \textbf{Mention:} ``Does this passage discuss the dimension in any way?'' A yes means the paper ``talks about the dimension as defined, anywhere in the paper: motivation, related work, the benchmark itself, results, limitations or future work all count.''
\item \textbf{Claim:} ``Does this passage state that the paper's own benchmark tests the dimension?'' A yes means the paper ``says the benchmark presented in this paper contains the dimension in its tasks, its environment, or its grader (for example, the agent must deal with it to complete tasks, the environment simulates it, or grading checks it).'' ``Discussion of other benchmarks or prior work, general motivation, limitations, or future work does not count.''
\item \textbf{Measure:} ``Does this passage report a result that measures the dimension separately?'' A yes means the paper ``reports a result of this paper's benchmark that isolates the dimension: a dedicated metric or score for it, an ablation that varies it, or a failure analysis that attributes errors to it.'' ``Results that only fold it into overall task success, or results about other things, do not count.''
\end{itemize}

\item \textbf{Subcomponents and the established-dimension rule.} For each claimed dimension, the model first records where it appears (task, environment or grader), the subcomponent values present, and, for task interdependence, authority structure and incentive structure, the other actor involved; only then does it decide whether the dimension is established. A claim counts only if a feature subcomponent is present, a scale subcomponent takes a non-baseline value, or the dimension fits the overall definition without any subcomponent (with a stated reason); task interdependence additionally requires another distinct actor. We also check this rule in code. For a measure, the model records whether the isolating result is quantitative (a metric, score or ablation) or qualitative (a failure analysis). Subcomponent values are recorded in the format listed in \hyperref[sec:E2]{section E2}.
\item \textbf{Quotations.} Every yes requires the model to give a verbatim quote of forty words or fewer from the paper to justify its label.
\item \textbf{Follow-up pass (Layers 2--4).} For each dimension the paper claims or measures, we make one model call per dimension, again with the full paper text and the items in \hyperref[sec:E3]{section E3}. These calls code the following:
\begin{itemize}
\item \textbf{Layer 2 (coordination requirements):} ``Does the paper claim the benchmark's tasks REQUIRE the agent to do this? Count it only if the task cannot be completed without it.''
\item \textbf{Layer 3 (governance mechanisms):} ``Does the paper claim the benchmark's environment SUPPLIES this mechanism?''
\item \textbf{Layer 4 (performance outcomes):} ``Does the paper claim the benchmark's grader CHECKS this outcome?''
\item Each is answered ``yes'', ``absent'' (the paper makes clear the benchmark does not have it) or ``not established'' (the paper does not say either way), and every yes again requires a verbatim quote.
\end{itemize}

\item \textbf{Final check.} We search the paper text for each quote, allowing for page breaks that interrupt it, and eliminate labels whose quote cannot be found.
\item \textbf{Consistency check.} We re-ran the Layer 1 tagging for 40 benchmarks with identical settings; the two runs agreed on 94 percent of claim tags.
\item \textbf{Development of the definitions.} The tagging definitions and notes were developed iteratively: we ran the tagging, reviewed the outputs against the concepts, and refined the definitions where the tags did not capture what we intended.
\end{enumerate}

\subsubsection*{E2. Layer 1 definitions: structure of work}\phantomsection\label{sec:E2}
\medskip\noindent\textit{Task interdependence}\par

\textbf{Definition:} The interdependency between the focal task and other activities, e.g. the output of one activity is the input to another.

\textbf{Subcomponents:} Pooled: Each activity makes a discrete contribution to the whole and is supported by it, without direct exchange between activities. Sequential: One activity's output becomes another activity's input, so the first must be completed before the second can proceed. Reciprocal: Each activity's output becomes an input for the other, so each poses a contingency for the other and must adjust as work proceeds.

\textbf{Notes on tagging:}

\begin{itemize}
\item The three forms are cumulative; tag the most demanding form present, not three separate yes/no variables.
\item Interdependence is between activities of different actors or departmental units within an organization; steps within one agent's own task do not count.
\item Task interdependence structure covers the order which arises because one activity's output is another's input. Where an order arises that states when a task's actions must occur (waiting for others, deadlines and windows, and interruptions), tag temporal structure instead.
\item Interdependence must be with other distinct actors in the environment. Sub-agents, child agents, and other agents that the focal agent creates or calls as tools are part of its own task and do not count.
\end{itemize}

{\raggedright Recorded as: form=pooled\textbar\allowbreak{}sequential\textbar\allowbreak{}reciprocal\textbar\allowbreak{}not\_established (the most demanding form present)\par}

\medskip\noindent\textit{Information structure}\par

\textbf{Definition:} The distribution, accessibility, and character of task-relevant knowledge across organizational actors and information systems, including whether acquiring or validating that knowledge depends on other actors' actions or authority.

\textbf{Subcomponents:} Location (scale): concentrated vs. distributed (held by other actors, and information systems, where acquisition and validation of that knowledge depends on other actors' actions or authority). Consistency (scale): consistent vs. contradictory. Codifiability (scale): tacit vs. explicit.

\textbf{Notes on tagging:}

\begin{itemize}
\item Documents that are supplied and can be accessed by the agent alone count as concentrated, even if there are a lot.
\item Knowledge that must be requested from a person is distributed (location). It is tacit only when it cannot readily be put into words, even by the person who holds it, and must instead be inferred from examples or practice.
\item Concentrated, consistent and explicit are baseline values: record them, but they do not establish the dimension alone.
\item Record all of the subcomponents (location, consistency, codifiability), with one of the two respective values chosen for each, e.g. choosing tacit for the codifiability subcomponent.
\end{itemize}

{\raggedright Recorded as: location=concentrated\textbar\allowbreak{}distributed\textbar\allowbreak{}not\_established; consistency=consistent\textbar\allowbreak{}contradictory\textbar\allowbreak{}not\_established; codifiability=tacit\textbar\allowbreak{}explicit\textbar\allowbreak{}not\_established\par}

\medskip\noindent\textit{Authority structure}\par

\textbf{Definition:} The authority to request, decide, approve, override and oblige.

\textbf{Subcomponents:} Formal authority (feature): the right to decide, derived from position or contract. Real authority (feature): effective control, in part from holding relevant knowledge.

\textbf{Notes on tagging:}

\begin{itemize}
\item Record formal or real authority separately, each as present or absent within the task.
\item Real authority applies when the effective control lies with an actor that is different from the formal holder. Formal authority can be present without real authority, while the presence of real authority also implies that formal authority is present. Both can therefore be present in the same task.
\item A written policy counts as formal authority when it instructs the role (what the actor may decide or grant on the organization's behalf), not when it only specifies the task.
\end{itemize}

{\raggedright Recorded as: formal=present\textbar\allowbreak{}absent; real=present\textbar\allowbreak{}absent\par}

\medskip\noindent\textit{Temporal structure}\par

\textbf{Definition:} The timing and order in which things must take place.

\textbf{Subcomponents:} Latency (feature): waiting time for input or other actor's response. Sequencing (feature): order of tasks across time. Time constraint (feature): deadlines, windows, interruptions.

\textbf{Notes on tagging:}

\begin{itemize}
\item Temporal structure covers when the task's actions must occur: waiting for others, deadlines and windows, and interruptions. Where an order arises only because one activity's output is another's input, tag task interdependence instead.
\item Time limits count when they are part of the task as presented to the agent (e.g., a deadline set by a client or manager). Limits on running the evaluation (wall-clock, compute, step budgets) do not.
\item Record the present subcomponents in the task, e.g. one task may have all three whereas another may only have 1.
\item Temporal sequencing counts when the order is instituted by the organization (such as other actors, schedules, policies, norms, or processes), even if the focal actor carries out all of the steps itself. A sequential order that arises only from the logic of the actor's own steps, rather than the organization, does not count.
\item Time constraints count when they limit when the focal actor must act in the task, such as a deadline, a time window to act in, or an interruption embedded in the task. Dates in the supplied material or information that the actor works with, such as checking dates within a period or understanding whether offers are still active on a date, do not count.
\end{itemize}

{\raggedright Recorded as: latency=present\textbar\allowbreak{}absent; sequencing=present\textbar\allowbreak{}absent; time\_constraint=present\textbar\allowbreak{}absent\par}

\medskip\noindent\textit{Incentive structure}\par

\textbf{Definition:} The alignment or divergence of actors' objectives and the consequences attached to their actions, including rewards, penalties, performance measures, and other mechanisms that shape behavior.

\textbf{Subcomponents:} Alignment (scale): aligned, divergent, or conflicting interests. Consequence-bearing (feature): whether the focal actor bears the consequences of their actions within the organization, such as rewards, penalties, or other mechanisms tied to performance measures.

\textbf{Notes on tagging:}

\begin{itemize}
\item Aligned interests are the baseline; only divergent or conflicting interests, or consequence-bearing, establish the dimension.
\item The benchmark's score does not count as a consequence in itself; consequences count when they exist within the task environment (e.g., budgets, earnings, penalties), either accruing during the task (e.g., a budget drawn down or built up, penalties) or received at its end (e.g., a payout for completed work), even if they also enter the final score. Consequences that only count toward the overall benchmark score, without being embedded in the task, do not. Record which consequences are present (rewards, penalties, and other mechanisms) and the performance measures they are tied to.
\item Record alignment or consequence-bearing separately, where one can be present without the other.
\item Interests count when another agent (even if simulated) in the task, inside or outside the organization (such as a customer), pursues objectives that differ or conflict with those of the focal actor.
\item Persistence or difficulty in behavior does not in itself count as divergent or conflicting objectives. The other actor must want a different outcome, not just be unclear, slow or repetitive in reaching the same one.
\end{itemize}

{\raggedright Recorded as: alignment=aligned\textbar\allowbreak{}divergent\textbar\allowbreak{}conflicting\textbar\allowbreak{}not\_established; consequence\_bearing=present\textbar\allowbreak{}absent; consequences=\textless{}which rewards/penalties/mechanisms and the performance measures they are tied to, or none\textgreater{}\par}

\medskip\noindent\textit{Environmental dynamism}\par

\textbf{Definition:} The stability of the environmental conditions relevant to the task while work is underway.

\textbf{Subcomponents:} Object of change (feature): requirements, roles, resources, or conditions that the task rests on. Predictability of change (scale): predictable vs. unpredictable.

\textbf{Notes on tagging:}

\begin{itemize}
\item How knowledge is distributed at the outset is information structure; change during execution is dynamism.
\item The objects of change can be present and vary independently of each other.
\item Any change in an object of change establishes the dimension, even when predictable; predictability is recorded but does not decide whether the dimension is present.
\item Change must not result from the focal actor's own actions. Changes to the actor itself, such as receiving a new task or a change in its role, count.
\end{itemize}

{\raggedright Recorded as: change=\textless{}one or more of requirements, roles, resources, conditions, or none\textgreater{}; predictability=predictable\textbar\allowbreak{}unpredictable\textbar\allowbreak{}not\_established\par}

\medskip\noindent\textit{Overall tagging notes}\par

\begin{itemize}
\item First record whether the benchmark captures the dimension according to the definition (including subcomponents), and then record which subcomponents are present.
\item Subcomponents are of two types: \emph{scale subcomponents} record one value, or ''not established''; \emph{feature subcomponents} record all that are present.
\item A dimension is established when a feature subcomponent is present or a scale subcomponent takes a non-baseline value. A dimension can also be present if it fits with the overall definition of the dimension without any of the subcomponents.
\end{itemize}

\subsubsection*{E3. Layers 2--4 items}\phantomsection\label{sec:E3}
Asked in the follow-up pass for each claimed or measured dimension. An item that combines several elements, such as information acquisition and reconciliation, is tagged as a single item.

\medskip\noindent\textit{Task interdependence}\par

\begin{itemize}
\item \textbf{Coordination requirement (Layer 2):} Consistency of outputs (pooled); Synchronization (sequential); Mutual adjustment (reciprocal). \emph{Managing interdependencies through actions that are consistent, synchronized or mutually adjustable as work proceeds.}
\item \textbf{Governance mechanism (Layer 3):} Rules and standards (pooled); Plans and schedules (sequential); Direct communication (reciprocal). \emph{Managing different forms of interdependencies through standardization, planning, and communication.}
\item \textbf{Performance outcome (Layer 4):} Consistent inputs (pooled); Timely and complete hand-offs (sequential); Successful adjustments (reciprocal). \emph{Successful navigation of dependencies, while satisfying required quality of focal output.}
\end{itemize}

\medskip\noindent\textit{Information structure}\par

\begin{itemize}
\item \textbf{Coordination requirement (Layer 2):} Information acquisition and reconciliation. \emph{Obtaining fragmented pieces of information, reconciling potentially conflicting views and eliciting tacit knowledge inputs.}
\item \textbf{Governance mechanism (Layer 3):} Information rules and standards. \emph{Structuring information distribution and codification through standard-setting and information systems.}
\item \textbf{Performance outcome (Layer 4):} Correct and complete information obtained; Contradictory information resolved. \emph{The acquisition and reconciliation of information, capturing its nuances to accurately represent the context.}
\end{itemize}

\medskip\noindent\textit{Authority structure}\par

\begin{itemize}
\item \textbf{Coordination requirement (Layer 2):} Routing and authorization. \emph{Routing decisions to actors with relevant decision rights, formally or in practice, and obtaining the necessary approval before proceeding.}
\item \textbf{Governance mechanism (Layer 3):} Formalized decision rights. \emph{Allocating decision rights which facilitate approvals, delegation, escalation.}
\item \textbf{Performance outcome (Layer 4):} Necessary approvals obtained; Proper organizational conduct. \emph{Decision rights, confidentiality, and task ownership respected.}
\end{itemize}

\medskip\noindent\textit{Temporal structure}\par

\begin{itemize}
\item \textbf{Coordination requirement (Layer 2):} Sequencing, waiting, and prioritization. \emph{Understanding the correct sequence in which tasks should be executed, waiting for inputs, and prioritizing certain tasks over others.}
\item \textbf{Governance mechanism (Layer 3):} Planning and scheduling. \emph{Formalizing the schedule in which tasks should take place, when deadlines should be met, and in which order the tasks are prioritized.}
\item \textbf{Performance outcome (Layer 4):} Activities performed in required order; Deadlines and time windows satisfied. \emph{Output produced in correct temporal order with respect for sequencing, deadlines, and interruptions.}
\end{itemize}

\medskip\noindent\textit{Incentive structure}\par

\begin{itemize}
\item \textbf{Coordination requirement (Layer 2):} Alignment and negotiation. \emph{Securing the aligned participation of actors whose interests may diverge, partly through formal and informal negotiation.}
\item \textbf{Governance mechanism (Layer 3):} Incentives and monitoring. \emph{Introducing incentives to align objectives of involved actors, alongside monitoring structures to mitigate agency problems.}
\item \textbf{Performance outcome (Layer 4):} Cooperation of actors achieved; Divergent aims recognized and managed. \emph{Objectives of the involved actors successfully aligned with the focal task.}
\end{itemize}

\medskip\noindent\textit{Environmental dynamism}\par

\begin{itemize}
\item \textbf{Coordination requirement (Layer 2):} Anticipation and adaptation. \emph{Anticipating and adapting to changes in environmental conditions relevant to the task.}
\item \textbf{Governance mechanism (Layer 3):} Dynamic planning and communication. \emph{Allowing for flexibility in tasks in order to continuously adapt, replan and communicate when environmental conditions change.}
\item \textbf{Performance outcome (Layer 4):} Plans and outputs adapted to emergent information; Changes in tasks successfully communicated. \emph{Output aligned with changes in environmental conditions, alongside successful communication to other involved actors.}
\end{itemize}

\subsubsection*{E4. Classifier prompt for selecting benchmark papers}\phantomsection\label{sec:E4}
\begin{promptblock}
You classify AI research papers from their title and abstract only. For each paper answer:

- is\_benchmark: true only if introducing a new benchmark, evaluation dataset, evaluation environment, arena or test suite for evaluating AI models (LLMs, VLMs, foundation models or AI agents) is a PRIMARY contribution of the paper (typically signalled by naming the benchmark and describing it as the contribution). False for papers whose main contribution is a method, model, framework, analysis or training dataset, even if they also build a small evaluation set; false for papers that only use existing benchmarks or study humans.

- evaluates\_agents: true if the benchmark evaluates AI agents that take actions over multiple steps in an environment (tools, web, OS, games, simulations, multi-turn interaction), false for single-turn question answering or classification.

- game\_based: true if the benchmark is built on games or game-like play (board, card, video, text adventure, social deduction, negotiation or economic games, game-theoretic interactions).

- organizational: true if the benchmark explicitly places the agent in a workplace or organizational setting, or with other actors whose roles, decisions, information or interests matter (colleagues, customers, managers, other agents with their own goals).

Judge only from the text given. If the abstract is missing, judge from the title and be conservative. Return one row per paper, using the paper's id exactly as given.

\end{promptblock}

\end{document}